\documentclass[12pt]{article}
\usepackage{graphicx}
\usepackage{tikz} % figures, diagrams, ..

\usepackage[textwidth=8em,textsize=small]{todonotes}
\usepackage{amsmath,amssymb} %amsthm,amsfonts
\usepackage{natbib}
\usepackage{dsfont}
\usepackage{booktabs}
\usepackage{bm}
\usepackage{hyperref}
\usepackage{xurl}
\usepackage{mathtools}
\usepackage[normalem]{ulem}

\newcommand{\mo}[1]{\textcolor{black}{#1}}

\usepackage{colortbl}
\usepackage{subfigure}
\usepackage{makecell}
\usepackage{lscape}

\usepackage{etoolbox}
\makeatletter
\patchcmd{\@makecaption}
  {\parbox}
  {\advance\@tempdima-\fontdimen2}
  {}{}
\makeatother 

\newcommand{\blind}{1} %blind version

\newtheorem{theorem}{Theorem}
\newtheorem{lemma}[theorem]{Lemma}
\newtheorem{proposition}[theorem]{Proposition}
\newtheorem{definition}[theorem]{Definition}
\newtheorem{remark}[theorem]{Remark}
\newtheorem{example}[theorem]{Example}

\newcommand{\NN}{\mathbb{N}}

\newcommand{\RR}{\mathbb{R}}

\newcommand{\Yu}{Y^{(u)}}
\newcommand{\Yv}{Y^{(v)}}

\newcommand{\Yun}{Y^{(u_n)}}
\newcommand{\Yvn}{Y^{(v_n)}}

\newcommand{\gu}{g^{(u)}}
\newcommand{\gv}{g^{(v)}}
\newcommand{\gun}{g^{(u_n)}}
\newcommand{\gvn}{g^{(v_n)}}

\newcommand{\Ru}{R^{(u)}}
\newcommand{\Run}{R^{(u_n)}}
\newcommand{\Ruv}{R^{(u,v)}}
\newcommand{\limeps}{\lim_{\substack{\mo{u \to \infty}\\ v/u \to \varepsilon}}}
\newcommand{\Rhatunvn}{\widehat R_n^{(u_n,v_n)}}
\newcommand{\Rhatuv}{\widehat R_n^{(u,v)}}
\newcommand{\Reps}{R_{\varepsilon}}

\newcommand{\cepsone}{c_{\varepsilon}^{(1)}}
\newcommand{\cepstwo}{c_{\varepsilon}^{(2)}}
\newcommand{\rreps}{r_{\varepsilon}} 
\newcommand{\qepsonetwo}{q_{\varepsilon}^{(12)}}
\newcommand{\qepstwoone}{q_{\varepsilon}^{(21)}}
\newcommand{\peps}{p_{\varepsilon}}

\newcommand{\one}{\mathds{1}}

\newcommand{\Rbar}{R}

\DeclareMathOperator{\EE}{\mathbb{E}}
\DeclareMathOperator{\PP}{\mathbb{P}}
\DeclareMathOperator{\Var}{Var}
\DeclareMathOperator{\Cov}{Cov}
\begin{document}

%%%%%%%%%%%%%%%%%%%%%%%%%%%%%%%%%%%%%%%%%%%%%%%%%%%%%%%%%%%%%%%%%%%%%%%%%%%%%%
\if1\blind
{
  \title{\bf Evaluation of binary classifiers for 
asymptotically dependent and independent extremes}
  \author{ Juliette Legrand$^{a,b}$, Philippe Naveau$^{b}$ and Marco Oesting$^c$ \thanks{
    Part of this  work was supported by funded by Deutsche Forschungsgemeinschaft (DFG, German Research Foundation) under Germany's Excellence Strategy - EXC 2075 - 390740016 and the Stuttgart Center for Simulation Science (SimTech).
 Another part was supported by European H2020 XAIDA (Grant agreement ID: 101003469) and the  French  Agence Nationale de la Recherche:  EXSTA (ANR-23-CE40-0009), the PEPR TRACCS programme under grant number (PC4 EXTENDING, ANR-22-EXTR-0005),  the PEPR   IRIMONT (France 2030 ANR-22-EXIR-0003) and the SHARE PEPR Maths-Vives (France 2030 ANR-24-EXMA-0008).
Naveau’s research work has also benefited  from the Geolearning research chair, a joint initiative of Mines Paris and the French National Institute for Agricultural Research (INRAE).
 }\hspace{.2cm}\\~\\
 $^a$Univ Brest, CNRS UMR 6205,\\ Laboratoire de Mathématiques de Bretagne Atlantique, France\\
    $^b$Laboratoire des sciences du climat et de l'environnement (EstimR), \\ Université Paris-Saclay, CNRS, CEA, UVSQ,
91191 Gif-sur-Yvette, 
France\\
   % Philippe Naveau \\
%     Laboratoire des sciences du climat et de l'environnement (EstimR), \\ Université Paris-Saclay, CNRS, CEA, UVSQ,
% 91191 Gif-sur-Yvette, 
% France\\
% and \\
%     Marco Oesting \\
    $^c$Stuttgart Center of Simulation Science \& Institute for Stochastics and \\Applications, University of Stuttgart, 
70563 Stuttgart, 
Germany}
\date{}
  \maketitle
} \fi

\if0\blind
{
  \bigskip
  \bigskip
  \bigskip
  \begin{center}
    {\LARGE\bf Evaluation of binary classifiers for asymptotically dependent and independent extremes}
\end{center}
  \medskip
} \fi

\bigskip
\begin{abstract}
Machine learning classification methods usually assume that all possible classes are sufficiently present within the training set. Due to their inherent rarities, extreme events are always under-represented and classifiers tailored for predicting extremes need to be carefully designed to handle this under-representation. 
In this paper, we address the question of how to assess and compare classifiers with respect to their capacity  to capture extreme occurrences. 
This is also related to the topic of scoring rules used in forecasting literature. In this context, we propose and study a risk function adapted to extremal classifiers. The inferential properties of our empirical risk estimator are derived under the framework of multivariate  regular variation and hidden regular variation. 
A simulation study compares different classifiers and indicates their performance with respect to our risk function. 
To conclude, we apply our framework to the analysis of extreme river discharges in the Danube river basin. 
The application compares different predictive algorithms and test their capacity at  forecasting river discharges from other river stations. 
% As a byproduct, 
%  we study  the special class of linear classifiers,  show that the optimization of our risk function leads to a consistent solution
%  and we identify the explanatory variables that contribute the most to extremal behavior.
 \end{abstract}

\noindent%
\vfill

\newpage
%\spacingset{1.9}
\section{Introduction}
\label{sec::intro papier marco}

In statistics, extreme events are generally 
defined as events with a very low probability of occurrence
\cite[see, e.g.,][]{resnick2003,Asadi2015}. 
For many applications in finance, environmental sciences and risk analysis, such rare  events can be viewed as the binary response of a complex system driven by a set of  explanatory variables \cite[see, e.g.,][]{Mhalla2020}.  
To illustrate the binary nature of extremes in practice, one can look at  the \href{https://www.weather.gov/lwx/WarningsDefined#Excessive%20Heat%20Watch}{warming definitions}  of the {US national weather service}. 
For example, 
ice accumulation above $1/4$ inch ($1$ cm)  implies  ice storm warnings or a heat index value above  110 degrees Fahrenheit (43 degree Celsius) leads to an excessive heat warning (east of the Blue Ridge).
In these two  cases, the forecast warming setup can be expressed as a binary imbalanced classifiers problem, \textit{i.e.}, one class label (extreme events) has a very low number of observations and the other category (non extreme events) is over represented 
\cite[see, e.g.,][]{He2013,Jalalzai2018}. 
Concerning health issues, the \href{https://www.eea.europa.eu/en/analysis/maps-and-charts/particulate-matter-pm10-annual-limit-value-for-the-protection-of-human-health-3}{air quality directive (2008/EC/50)} has set two limit values for particulate matter (PM10) for the protection of human health in Europe: the PM10 daily mean value may not exceed 50 micrograms per cubic metre  more than 35 times in a year and the PM10 annual mean value may not exceed 40 micrograms per cubic metre. 
Such examples show that there is a   great societal and economical interest for  risk managers to assess how statistical models  can predict  occurrences of such large events, \textit{i.e.}, to compare  binary classifiers that aim at  predicting a sequence of $\{-1,+1\}$ where $+1$ corresponds to a rare event. 
The frequency of $+1$ usually decreases  with the level of risk.
This notion of risk level can be easily illustrated by   hurricanes forecasts 
from the warming system of the \href{https://www.weather.gov/hgx/tropical_scale}{US national weather service}. Hurricanes  are ranked into five categories according to their wind speeds, the so-called Saffir-Simpson Hurricane Wind Scale, where the highest category corresponds to  wind speeds above 156 mph (251 km/h), and the second largest to the range 130-156 mph  (209-251 km/h). Counting events above the two predetermined wind speed thresholds leads to a natural ranking between warning categories. 
To analyze such cases, we  always assume  in this work that  a data set like Table \ref{table::ingredients} 
is given 
to us. For hurricane categories 4  and 5,  $v=130$ mph and $u=156$ mph will be plugged 
in Table~\ref{table::ingredients}.
This type of setup is general and contains all examples listed above. 
We also want to emphasize that, our goal is not to construct a new binary classifier to  issue  warnings, but rather to propose new risk functions that compare  existing binary classifiers that have been trained on    binary samples corresponding to these rare event occurrences. 
Here, we authorize the data scientist to change his/her classifier 
according to  the chosen  risk, say the   hurricane category, but we also require   classifiers under scrutiny  to be  risk consistent. 
This simply means that a forecaster cannot provide a category 5 hurricane forecast and, under the same condition, the absence of a category 4, see  Definition 1 in Section  \ref{sec: risk functions} for mathematical details. 

%%%%%%%%%%%%%%%%%%%%%%%%%%%%%%%%%%%%%%%%%%%%%%%%%%%%%%%%%%
\begin{table}[h]
\caption[ingredients ]{\label{table::ingredients} Ingredients to compare  classifiers for extremes in  this study (numbers in bold corresponds to misses; either false positives or false negatives). 
Extreme binary events change with the threshold choice. Here the threshold $u$ is greater than $v$. 
For example,  \href{https://www.weather.gov/hgx/tropical_scale}{hurricane   events of  categories 4 and 5} are defined from  two predetermined wind speeds  above $u=156$ mph (251 km/h), and above $v=130$ mph  (209 km/h), respectively.}
\vspace{3mm}
%\centering
\raggedright 
%\toprule
   \hspace{25mm} Observed events \hspace{10mm}  Forecasted classifier outputs \\
   \hspace{25mm}  above  threshold \hspace{10mm} for threshold\\
% \midrule 
 \begin{tabular}{c}
 % \toprule
Label \\ 
  \midrule
  1 \\ 
  2 \\ 
  3  \\ 
  %-1  & -1 \\ 
  $\vdots$  \\ 
  $n-2$ \\ 
  $n-1$ \\ 
  $n$ \\ 
  %X13 & \textcolor{red}{\textbf{0.30}} \\ 
  \bottomrule
  \end{tabular}
  \hspace{4em} 
\begin{tabular}{cc}
 % \toprule
 $v$  &  $u$ \\ 
  \midrule
  -1 & -1 \\ 
  +1 & -1 \\ 
  +1 & +1 \\ 
  %-1  & -1 \\ 
  $\vdots$ & $\vdots$ \\  
  -1 & -1 \\ 
  +1 & +1 \\ 
  -1 & -1 \\ 
  %X13 & \textcolor{red}{\textbf{0.30}} \\ 
  \bottomrule
  \end{tabular}
  \hspace{5em} 
\begin{tabular}{rr}
 % \toprule
 $v$  & \;\; $u$ \\ 
  \midrule 
 {\bf +1} & -1 \\ 
  +1 & -1 \\ 
  {\bf -1} & {\bf -1} \\ 
  $\vdots$ & $\vdots$ \\ 
   -1  & -1 \\ 
 +1 & {\bf -1} \\ 
  -1  & -1 \\ 
  %X13 & \textcolor{red}{\textbf{0.30}} \\ 
   \bottomrule
\end{tabular}
\end{table}
%%%%%%%%%%%%%%%%%%%%%%%%%%%%%%%%%%%%%%%%
In this context, one main question is to determine which risk functions can insure coherent rankings between classifiers,  and also  to identify under which theoretical umbrella probability results and inferential guarantees can be given.
The later point can be addressed by leveraging multivariate extreme  value theory (MEVT)  that provides  mathematically sound dependence structures \cite[see, e.g.,][]{Coles1999} for modeling complex joint tail behaviors across multiple variables.

Concerning binary classifiers, it is important to understand why basic loss functions are inefficient at  comparing classifiers for extremes. 
To see this, we can revisit 
the classical expected loss that  simply counts the number of times that a given classifier provides a wrong answer. 
For example, the expected loss applied to  the ``always optimistic"  classifier that never forecasts $+1$  appears to improve  as $u$ increases (fewer and fewer extremes are observed), and it even  goes to zero for very large values of $u$. 
So, never predicting an extreme appears to be a optimal strategy with respect to the classical expected loss.

To avoid this undesirable feature, one natural idea is to re-scale the expected loss by the probability of observing an extreme.
In this case, the rescaled associated risk now goes towards one as $u$ gets large. 
But, this does not fix the imbalance nature in assessing extreme classifiers. To explain this, one can introduce 
another trivial case:  the ``crying wolf" classifier 
who always predicts an extreme, see also the forecaster's dilemma \cite[e.g.][]{Lerch2017}. 
In this case, one can show that the previous rescaling trick does not work as  the rescaled loss function  goes towards infinity  as $u$ gets large. 
This wrongly indicates that  the ``crying wolf" classifier  is much worse than the overly optimistic one. 
Both of them are  unreasonable in practice and there is no reason to strongly favour one over the other one. 
For this reason,  we propose and study a  new class of loss functions, see Equation \eqref{eq: sets}, that has the advantage of handling such imbalanced cases and be in compliance with MEVT. 
We will show that our new loss function has  strong connections with existing scores and can be viewed as an  extension of some classical loss functions.   
In particular, the critical success index, also called  the threat score in the weather literature \cite[e.g.][]{Schaefer1990} is closely linked to this risk function. 
This index  computes the total number of correct event forecasts (hits) divided by the total number of  forecasts plus the number of misses (hits + false alarms + misses).
In the context of rare events forecasts, \cite{Stephenson2008}  highlighted some advantages and drawbacks of various risk functions, including  the critical success index.
These authors linked forecast scoring rules  with two dependence indices used in EVT \cite[see][]{Coles1999}.
The first one is the tail coefficient, called $\chi$ in EVT,  that provides the probability of true positive event given a extreme event whenever the forecast can be assumed to be calibrated, \textit{i.e.}\  observations and forecasts follow the same marginal  distributions.
In MEVT, a strictly positive $\chi$ indicates {\it asymptotic dependence}, compare also Definition \ref{def: tail equivalent} and Remark \ref{rem: tail equívalent}.
The second index 
captures some second order extremal dependence information, called  \emph{asymptotic independence} in MEVT. It is linked to the tail dependence coefficient $\eta$ through the relationship $2\eta-1$ (see Equation \eqref{eq: ramos} for the definition of $\eta$).
\cite{Stephenson2008} advocated the use of  this index and coined the term: the extreme dependency score. 
Later on, \cite{Ferro2011} proposed two  different scores and studied their properties.  
But their link with the concept of asymptotic  independence was not clear and the convergence results of their estimators were not fully developed. 
Hence, there is also  a mathematical interest   to study loss functions for  the asymptotic independent case.

In the machine learning literature,  there is a vast body of work about imbalanced data 
\cite[see, e.g.,][]{Hilario2018}. 
For example, the so-called  functions of precision and recall are two metrics well-used to score binary classifiers in learning research, see, e.g., \cite{He2013} and the $F$-score. 
Still, the review paper by \cite{Haixiang2017} did not mention any imbalanced method based on EVT, even though some studies have proposed binary regression models for imbalanced data based on univariate EVT distributions \citep[e.g.][]{Wang2010}.
Nonetheless, to our knowledge, very few theoretical links have been made to bridge  imbalanced learning and  MEVT.  
One noticeable exception is \cite{Jalalzai2018} who  worked on  binary classifiers for extremes under the regular variation hypothesis \cite[see, e.g.,][]{resnick2003}. 
But they differ from our work on two aspects. 
First, their object of interest was  extremes defined as rare events coming from binary classifiers. 
In our case, we consider extremes as rare events from either the responses or the binary classifier outputs. 
The second difference with our work is that, although they considered in details many mathematical aspects under the  asymptotic dependence case, they did not explore the asymptotic independence situation. In this study, one part of our results is based on the  concept of hidden regular variation \cite[see, e.g.,][]{Ledford96,heffernan2005,Ferro2007}. 
In particular, we take advantage of the model of \cite{Ramos2009} to derive the asymptotic properties of our estimators. 
Table \ref{table::review_risks} in Appendix \ref{sec:review_risks} summarizes the main features of some risk measures used to compare binary classifiers for extremes, see also Table I in \cite{Stephenson2008} for a score comparison.

Our paper is organized as follows. 
In Section \ref{sec: risk functions}, we propose and study a  risk function that can handle  both the asymptotic dependence and asymptotic independence cases. 
Estimators are also constructed and their asymptotic properties derived. Section \ref{sec::simulations} focuses on a simulation example that highlights the difficulty to compare common classifiers in the case of asymptotic independence. 
In Section \ref{sec: Danube river discharges}, we revisit the well studied example of the Danube river application and see how the choice of the metric can change the ranking of classifiers. 
Note that, besides the proofs of all propositions, the appendix   contains a statistical test for classifiers performances, see Appendix \ref{appendix::test}, but also addresses the questions of how to optimize the linear classifier for extremes and how to choose the relevant features, see Appendix \ref{sec:lin_class}.

\section{Risk functions for extremal binary classifiers}
\label{sec: risk functions}

 As already mentioned in the introduction, we consider binary response variables $\Yu \in \{-1,+1\}$ depending on some index $u$ which we interpret as level of rarity of events. Thus, an increasing level $u$ means that the events considered to be extreme are becoming more and more rare.
For example,
two events $\{\Yu = +1\}$ and $\{\Yv = +1\}$ in Table \ref{table::ingredients} are nested in the sense of  the ranking between the columns named $u$ and $v$. Concerning binary classifiers, the same level-dependent property appears as a natural requirement to compare extremal classifiers.  To mathematically formalize this condition, we denote 
  $\gu(\bm X)$ the threshold-dependent classifier based on the 
  covariate set \footnote{As stated in the introduction, only a table like Table \ref{table::ingredients} is needed to compute the scores studied in this work.  The notation $\bm X$ simply highlights how classifiers can be built. } $\bm X$. The level-dependent condition  can be  given then. 
\begin{definition}\label{def: level-dependent}
 A collection of random variables $Y=\{\Yu\}_{u \geq 0}$ 
 is said to a level-dependent binary response  
 whenever 
$$ \{ \Yu = +1 \} \subseteq \{ \Yv = +1\} \quad \text{for } u \geq v.$$
Similarly, a level-dependent binary classifier $g=\{\gu\}_{u \geq 0}$ is  a set of measurable functions $\gu : \RR^d \to \{-1,+1\}$
that are ordered in the sense that
$$ \{\gu(\bm X) = +1\} \subseteq \{\gv(\bm X)=+1\} \quad
\text{for } u \geq v, $$
 meaning that the event $\{\gu(\bm X) = +1\}$ implies the event $\{\gv(\bm X) = +1\}$ whenever $u\geq v$.
\end{definition}

\begin{example} \label{ex:latent}
 A typical example of a level-dependent binary response is a response via a positive latent variable $Y^*$ with upper end point $\infty$:
$$ \Yu = \begin{cases}
   +1, & \quad \text{if } Y^* > u,\\
   -1, & \quad \text{if } Y^* \leq u.    
\end{cases}$$
In the example of Hurricane classification mentioned in the introduction, the actual wind speed can be interpreted as such a (latent) variable where $u$ corresponds to one of the thresholds used for the definition of Hurricane classes.

For such a response, $u \geq v$ automatically implies that 
$$ \{ \Yu = +1 \}  = \left\{ Y^* > u \right\} 
\subseteq \left\{ Y^* > v \right\}  = \{ \Yv = +1\}.  $$
Given this property of the response variable, it is reasonable to assume the same property for binary classifiers.
\end{example}

Moreover, we set the range of the level $u$ to $[0,\infty)$ with $\PP(Y^{(0)}=+1)=1$, \textit{i.e.}\ the response is always considered as extreme at the minimal level $u=0$, while $\PP(Y^{(u)}=+1) \to 0$ as $u \to \infty$. 
Note that fixing the  threshold range to $[0,\infty)$  is arbitrary and it could be replaced if  the data range  was different. 
This convention simplifies notations (avoid integral transform) and eases the interpretation of extremes. 
\medskip

The fourth and fifth  columns of Table \ref{table::ingredients} illustrates the nested ordering of $g$.
Without loss of generality, we have assumed in this definition that all the classifiers to be considered depend on the same vector of input variables $\bm X \in \RR^d$.

\subsection{Overview over risk functions and extremal dependence}

 As outlined in the introduction, a reasonable risk function should be based on the classical classification loss $\one\{\gu(\bm X) \neq \Yu\}$ and should, at least asymptotically, penalize the 
two trivial estimators ``always optimistic'' and ``crying wolf'' equally. A possible choice, satisfying these two conditions is the weighted classification loss function
$$ \ell_u(g;(\bm x,y)) = \frac{1}{\PP(\Yu =+1  \text{ or } \gu(\bm X)=+1)} \one\{\gu(\bm x) \neq y\}
$$
and the associated expected risk
\begin{equation}\label{eq:Ru}
    \Ru(g) = \EE(\ell_u(g;(\bm X,\Yu))) = 
    \frac{\PP(\gu(\bm X) \neq \Yu)}{\PP(\Yu=+1 \text{ or } \gu(\bm X) =+1)}.
\end{equation}
 Since \begin{align*}\{\gu(\bm X) \neq \Yu\} &= \{(\gu(\bm X) = -1 \text{ and } \Yu=+1) \text{ or } (\gu(\bm X) = +1 \text{ and } \Yu=-1)\} \\
&\subseteq \{\Yu=+1 \text{ or } \gu(\bm X)=+1\}, \end{align*}
necessarily, $\Ru(g) \in [0,1]$, with $\Ru(g)=0$ indicating a perfect classifier and $\Ru(g)=1$ a bad one.
In particular, the ``always optimistic'' classifier $\gu \equiv -1$ possesses unit risk with $\Ru(g) =1$ 
at each level $u >0$.
Similarly, the risk of the ``crying wolf"  classifier $\gu \equiv +1$ is then equal to  
$\Ru(g) = {\color{black}\PP( \Yu =-1)}$. It  therefore converges to one as $u \to \infty$.
{\color{black} These two naive classifiers are, hence, equally  disqualified by $\Ru(g)$ for large $u$.}
This {\color{black} upper} unit value of $\Ru(g)$ provides a clear benchmark to assess any other {\color{black} level-dependent} binary classifiers $g$ satisfying the existence of the limit  $ R(g) =  \lim_{u \to \infty} \Ru(g).$
We call such classifiers \emph{extremal}  and $R(g)$ the \emph{extremal risk}, which has to satisfy $R(g) \in [0,1]$ by definition.
Then, $1-\Ru(g)$ can be understood  as a critical success index for extremes, \cite[e.g.][]{Schaefer1990}.

{\color{black} As will be described in Section \ref{subsec:extremalrisk}, $\Ru(g)$ can be calculated from the three probabilities  $\PP(\Yu=+1)$, $\PP(\gu(\bm X)=+1)$ and $\PP(\Yu=+1, \gu(\bm X)=+1)$. Perceiving these three expressions as functions of $u \in [0,\infty)$, they are all non-increasing and going from $1$ to $0$ and therefore have similar properties as tail functions of random variables. This allows us to define terms that are well-known from extreme value theory when considering tails of continuous random variables.
For instance, two random variables are called tail equivalent if both corresponding tail functions, \textit{i.e.}\ the probabilities of being extreme, decay at the same rate. Analogously, we could consider the two probabilities of observing an extreme event and classifying an event as extreme as the level of extremity tends to infinity. Similarly, the terms of asymptotic dependence and asymptotic independence focus on the probability of two random variables being jointly extreme, while we could take the probability of events being both observed and classified as an extreme.

\begin{definition}\label{def: tail equivalent}
Let $Y$ be a level-dependent binary response 
and $g$ be a level-dependent binary classifier, see Definition \ref{def: level-dependent}.
\begin{itemize}
    \item[1.] $Y$ and $g$ are called tail comparable if the limit
    $$ c(g) := \lim_{u \to \infty} \frac{\PP(\gu(\bm X)=+1)}{\PP(\Yu=+1)} \in [0,\infty] $$
    exists. They are called tail equivalent if
    $c(g) \in (0,\infty)$.
    Analogously, we say that $g$ has a lighter tail than $Y$ if $c(g)=0$, while $g$ is said to possess a heavier tail than $Y$ if $c(g)=\infty$.
    \item[2.] Assume that $Y$ and $g$ are tail equivalent. Then, we call them asymptotically dependent if
    $$ \chi^*(g) := \lim_{u \to \infty} \PP(\gu(\bm X)=+1 \mid \Yu=+1) > 0,$$
    while we call them asymptotically independent if
    $$ \lim_{u \to \infty} \PP(\gu(\bm X)=+1 \mid \Yu=+1) = \lim_{u \to \infty} \PP(\Yu=+1 \mid \gu(X)=+1) = 0.$$
\end{itemize}
\end{definition}
Note that tail equivalence of $Y$ and $g$ implies that the limits of two conditional probabilities $\PP(\gu(\bm X)=+1 \mid \Yu=+1)$ and $\PP(\Yu=+1 \mid \gu(\bm X)=+1)$ as $u \to\infty$ just differ by the factor $c(g)$. In particular, 
$Y$ and $g$ are asymptotically independent if and only if $\chi^*(g)=0$. If $c(g)=1$, $\chi^*$ is very closely related to the tail dependence coefficient $\chi$ of two random variables \cite[see][]{Coles1999} that is usually used to define asymptotic dependence and independence.

\begin{remark} \label{rem: tail equívalent}
    Note that the notions of tail equivalence and asymptotic (in-)dependence in Definition \ref{def: tail equivalent} are strongly motivated where, similarly to Example \ref{ex:latent}, there are unobservable latent variables $Y^*$ and $g^*(\bm X)$ such that
    $$ \Yu = + 1 \quad \iff \quad Y^* > u$$
    and 
    $$ \gu(\bm X) = + 1 \quad \iff \quad g^*(\bm X) > u. $$
    In this case, $Y$ and $g$ are tail equivalent in the sense of Definition \ref{def: tail equivalent} if and only if $Y^*$ and $g^*(\bm X)$ are tail equivalent in the usual sense. Analogously, asymptotic (in-)dependence of $Y$ and $g$ in the sense of Definition \ref{def: tail equivalent}  nearly correspond to the usual sense of asymptotic (in-)dependence of $Y^*$ and $g^*(\bm X)$ (remark: here we do not require the margins of $Y^*$ and $g^*(\bm X)$ to be identical, which differs from the usual definition of asymptotic (in-)dependence in EVT). 
\end{remark}

In Section \ref{subsec:extremalrisk}, we will demonstrate that both tail equivalence and asymptotic dependence of $Y$ and $g$ are necessary conditions to obtain an extremal risk $R(g)<1$, \textit{i.e.}\ a better performance than that of the naive classifiers. While a classifier $g$ can often be modified to become tail equivalent by some appropriate transformation $\varphi$ of the level $u$ to be considered, \textit{i.e.}, we compare $\Yu$ with $g^{(\varphi(u))}$ instead of $\gu$, there are situations where no asymptotic dependent classifier exists, but all tail equivalent classifiers are asymptotically independent, e.g., if the covariate vector $\bm X$ does not comprise the main drivers for extremes. Even in this case, however, it is still of interest to compare different asymptotically independent classifiers as the covariate vector $\bm X$ might contain at least some information on the extremes. Thus, we need to modify the risk function under consideration.

To this end, we note that, in the case of asymptotic independence, we typically face the following situation where the probability mass %of the measurable set 
$B_u=\mathbb{P}\left(\gu(\bm X) = \Yu=+1\right)$ becomes negligible compared to the sum of probability masses 
$A_u + C_u=\mathbb{P}\left(\gu(\bm X)=+1 \right)+\mathbb{P}\left( \Yu = +1 \right)$,  as $u$ gets large. 
The top graph in Figure \ref{fig: sets} schematically displays the measurable sets over which the probabilities
$A_u$, $B_u$ and $C_u$ are computed. 
When  $B_u=o(A_u + C_u)$, 
%To solve this issue and assess  if  
the classifier $\gu$ needs to squeeze some second order information from $\bm X$, 
i.e. \ 
%one needs  
to zoom into a smaller set over which $A_u + C_u$ is estimated by reducing its size. 
This is possible by introducing a moderate threshold $v$ smaller than $u$. 
The bottom right panel of Figure \ref{fig: sets} displays the new sets, and their corresponding probabilities, that will be used to compute our new risk function $\Ruv$.}
%i.e. the probability  $\PP(\gu(\bm X)=\Yu=+1 \mid \gv(\bm X)=\Yv=+1)$ may not necessarily go to zero when $u>v$ get large. 
%

\begin{figure}[!h]
    \centering
\begin{tikzpicture}
% style des nœuds
\tikzstyle{debutfin}=[ellipse,draw,text=red]
\tikzstyle{instruct}=[rectangle,draw,fill=yellow!50]
\tikzstyle{test}=[diamond, aspect=2.5,thick, draw=blue,fill=yellow!50,text=blue]
\tikzstyle{es}=[rectangle,draw,rounded corners=4pt,fill=gray!20]
\tikzstyle{AD}=[rectangle,draw,rounded corners=4pt,fill=blue!5]
\tikzstyle{AI}=[rectangle,draw,rounded corners=4pt,fill=red!5]
% style des flèches
\tikzstyle{suite}=[->,>=stealth’,thick,rounded corners=4pt]
% placement des nœuds
%\node[debutfin] (debut) at (-2,5) {Début};
% GRAPH
\draw[color=gray,->,>=stealth] (0-2,0+5) -- (4.5-2,0+5);
\draw[color=gray,->,>=stealth] (0-2,0+5) -- (0-2,4.5+5);
\filldraw[color=gray!20] (3-2,3+5) -- (4-2,3+5) -- (4-2,4+5) -- (3-2,4+5) -- cycle; % B 
\filldraw[color=gray!50] (0-2,3+5) -- (3-2,3+5) -- (3-2,4+5) -- (0-2,4+5) -- cycle; % A
\filldraw[color=gray!50] (3-2,0+5) -- (4-2,0+5) -- (4-2,3+5) -- (3-2,3+5) -- cycle; % C
\draw (1-2,3.5+5) node {$A_u$};
\draw (3-2+.5,3.5+5) node {$B_u$};
\draw (3-2+.5,1+5) node {$C_u$};
\draw (-0.5-2,3+5) node {$u$};
\draw (3-2,-0.2+5) node {$u$};

% TEST CONDITION
\node[es] (lire) at (0,4) {$B_u=o(A_u + C_u)$ for large $u$?};
\draw[color=red!40,double,->,>=stealth] (0+1,4-.35) -- (0+3,1+1.5);
\draw[color=blue!40,double,->,>=stealth] (0-1,4-.35) -- (0-3,1+1.5);
\draw (0+3+0.25,3) node {Yes};
\draw (0-3-0.25,3) node {No};
\node[AD] (afficher) at (0-5,2) {Asymptotic dependence};
\draw[color=blue!40,double,->,>=stealth] (0-5,1.5) -- (0-5,.5);
%\draw (0-3,-2) node {$\Ruv=\frac{A_{u,v}+C_{u,v}}{A_{u,v}+C_{u,v}+B_{u,v}}$};
\node[AD] (afficher) at (0-5,-0) {$\Ru=\frac{A_u+C_u}{A_u+C_u+B_u}$};

\node[AI] (afficher) at (0+4,2) {Asymptotic independence};
\draw[color=red!40,double,->,>=stealth] (0+4,1.5) -- (0+4,.5);

% SECOND GRAPH
\draw (3-0.5-1,3-4.5) node {$u$};
\draw (3+3-1,-0.5-4.5) node {$u$};
\draw (3-0.5-1,2-4.5) node {$v$};
\draw (3+3-2,-0.5-4.5) node {$v$};
\filldraw[color=gray!20] (3+3-1,3-4.5) -- (3+4-1,3-4.5) -- (3+4-1,4-4.5) -- (3+3-1,4-4.5) -- cycle; % B
\filldraw[color=gray!50] (3+0-1+2,3-4.5) -- (3+3-1,3-4.5) -- (3+3-1,4-4.5) -- (3+0-1+2,4-4.5) -- cycle; % new A
\filldraw[color=gray!50] (3+3-1,0+2-4.5) -- (3+4-1,0+2-4.5) -- (3+4-1,3-4.5) -- (3+3-1,3-4.5) -- cycle; % new B 
\draw (3+1+8.5-8,3.5-4.5) node {$A_{u,v}$};
\draw (3+3.5-1,3.5-4.5) node {$B_{u,v}$};
\draw (3+3.5-1,1+1.5-4.5) node {$C_{u,v}$};
%\draw (3+2.5-1,-1.5-4.5) node {$\Ruv=\frac{A_{u,v}+C_{u,v}}{A_{u,v}+C_{u,v}+B_{u,v}}$};
\draw[color=gray,,->,>=stealth,dashed] (3+3-2,2-4.5) -- (3+4-0.5,2-4.5);
\draw[color=gray,,->,>=stealth,dashed] (3+3-2,2-4.5) -- (3+3-2,4.5-4.5);
\draw[color=gray,->,>=stealth] (3+0-1,0-4.5) -- (3+4-0.5,0-4.5);
\draw[color=gray,->,>=stealth] (3+0-1,0-4.5) -- (3+0-1,4.5-4.5);

%\draw[color=black,->,>=stealth] (0+3,-6.5) -- (0+3,-7.5);
\draw[color=red!40,double,->,>=stealth] (1,-4) -- (0-.5,-4);
%\draw (0+3,-8) node {$\Ruv=\frac{A_{u,v}+C_{u,v}}{A_{u,v}+C_{u,v}+B_{u,v}}$};
\node[AI] (lire) at (0-3,-4) {$\Ruv=\frac{A_{u,v}+C_{u,v}}{A_{u,v}+C_{u,v}+B_{u,v}}$};
\end{tikzpicture}
\caption{Schematic representations of the ratio $\Ru$ defined by 
\eqref{eq:Ru} and $\Ruv$ defined by \eqref{eq:Ruv}. The asymptotic dependent (resp. independent) case occurs when the probability mass $B_u$ is comparable (resp. negligible) to $A_u + C_u$, as $u$ gets large. 
By removing mass below $v$, the probability mass $B_{u,v}$ can remain comparable to  $A_{u,v} + C_{u,v}$ in the asymptotic  independent case.}
\label{fig: sets}
\end{figure}

 In other words, we restrict our attention to the event
$\{\gv(\bm X)=+1, \Yv=+1\}$ and evaluate the risk $\Ru$ on this smaller set. This leads to the \emph{conditional risk function}
    \begin{equation} \label{eq:Ruv}
    \Ruv(g) =  \frac{\PP(\gu(\bm X) \neq \Yu \mid \gv(\bm X)=\Yv=+1)}
                    {\PP(\gu(\bm X)=+1 \text{ or } \Yu = +1 \mid \gv(\bm X)=\Yv=+1)}, \quad u \geq v \geq 0.
    \end{equation}
As we assumed that $\PP(g^{(0)}(\bm X)=+1) = \PP(Y^{(0)}=+1) = 1$, we can see that the definition of the conditional risk function provides a generalization of the risk function $\Ru$ defined in \eqref{eq:Ru} with $\Ru(g) = R^{(u,0)}(g)$. Similar to the extremal risk, we can also consider some  extremal limit of the conditional risk by letting $u \to \infty$ and $v \to \infty$ at the same rate. More precisely, for every $\varepsilon \in (0,1)$, we define the \emph{extremal conditional risk} $\Reps$ by
\begin{equation} \label{eq:reps}
 \Reps(g) = \limeps \Ruv(g).
\end{equation}

In the following sections, we will first investigate the original risk function $\Ru(g) = R^{(u,0)}(g)$ defined in \eqref{eq:Ru} and study its properties for asymptotically dependent and \textcolor{black}{asymptotically} independent classifiers, before turning to more general conditional risk $\Ruv(g)$ from \eqref{eq:Ruv}. The following
lemma collects several results that will form the basis for our further analysis.

\begin{lemma}\label{lemma: sets}
Let $g$ be a level-dependent binary classifier with $g^{(0)}(\cdot) \equiv +1$ and let $u \geq v \geq 0$.  
Then, the conditional risk $\Ruv(g)$ can be written as 
\begin{align}\label{eq: sets}
\Ruv(g)
={}& 1 - \bigg[ 
      \frac{1}{\PP(\Yu=+1 \mid \gu(\bm X)=+1, \Yv = +1)} \nonumber \\
      & \qquad 
    + \frac{1}{\PP(\gu(\bm X)=+1\mid  \gv(\bm X)=+1, \Yu = +1)} - 1 \bigg]^{-1},
\end{align}

In addition, we have the three following properties for $R$:
\begin{enumerate}
    \item[(a)] For all $u \geq 0$, the conditional risk $\Ruv(g)$ is non-increasing in $v$ with $R^{(u,u)}(g)=0$.
    \item[(b)] Let  $h$ be another level-dependent binary classifier.\\
     If $\{ \gu(\bm X) = +1\} = \{h^{(u)}(\bm X) = +1\}$ and $\{ \gv(\bm X) = +1\} \subset  \{h^{(v)}(\bm X) = +1\}$
    for some $v < u$, then 
     $
    \Ruv(g) \leq  \Ruv(h). 
    $
    \item[(c)] Let $a,b$ be some positive constants and $c_{v,v'}$ be some positive function, such that, for any $v, v' \in (0,u]$,
\begin{equation}\label{eq: mixing}
    \PP(\gv(\bm X)=+1, Y^{(v')}=+1) = c_{v,v'} \left(\PP(\gv(\bm X)=+1)\right)^a \left(\PP(Y^{(v')}=+1)\right)^b.
\end{equation}
Then
\begin{align}\label{eq: mixing cons}
    \Ruv(g) ={}& 1- \bigg[ \frac{c_{v,u}}{c_{u,u}} \frac{1}{(\PP(\gu(\bm X)=+1 \mid \gv(\bm X)=+1))^a} \nonumber \\
      & \qquad  +  \frac{c_{u,v}}{c_{u,u}} \frac{1}{(\PP(\Yu=+1 \mid \Yv=+1))^b)} - 1 
    \bigg]^{-1}.
\end{align}
\end{enumerate}
\end{lemma}
We deduce from Equation \eqref{eq: sets} that $\Ruv(g)=0$ if and only if 
$$\PP( \Yu=+1 \mid \gu(\bm X) = \Yv=+1)= \PP( \gu(\bm X) = +1 \mid \gv(\bm X) = \Yu=+1)=1.$$
The second property of this lemma indicates that, for a given pair of thresholds $u \geq v \geq 0$,  the risk function becomes smaller if the function $\gv$ becomes as small as possible. 

Equation \eqref{eq: mixing} can be viewed as a mixing condition that leads to a simple  expression of $\Ruv(g)$ based on probabilities of disjoint events.

\subsection{The extremal risk $R$} \label{subsec:extremalrisk}

In this section, we will further investigate the original risk function $\Ru(g) = R^{(u,0)}(g)$ defined in \eqref{eq:Ru}. Applying Lemma \ref{lemma: sets} with $v=0$ and using the fact that
$ \PP(Y^{(0)}=+1) = \PP(g^{(0)}(\bm X)=+1) = 1$,
we obtain
$$
\Ru(g)  = 1- \left[ 
\frac{1}{\PP(  \Yu = +1 \mid \gu(\bm X) = +1 )} + \frac{1}{\PP( \gu(\bm X) = +1 \mid  \Yu=+1  )} -1 
    \right]^{-1}.
$$
This leads to the expression
\begin{align}\label{eq:ru-rewritten}
    \Ru(g) 
    &{}={} 1 -   \frac{\PP(\gu(\bm X) = + 1 \mid \Yu = +1 )} { 1- \PP(\gu(\bm X) = +1 \mid \Yu = +1 ) + \PP(\gu(\bm X) = +1)/\PP(\Yu=+1)}. 
\end{align}
{\color{black} We can use this representation to analyze the effect of tail equivalence of $Y$ and $g$ as defined in Definition \ref{def: tail equivalent}:
\begin{itemize}
    \item If $g$ has a (partially) lighter tail than $Y$ in the sense that $ \liminf_{u \to \infty} \PP(\gu(\bm X) = +1) / \PP(\Yu = +1) = 0$, we also have  $\liminf_{u \to \infty} \PP(\gu(\bm X) = +1 \mid \Yu=+1 ) = 0$, and consequently, by Equation \eqref{eq:ru-rewritten}, 
    $$ \limsup_{u \to \infty} \Ru(g) =  1. $$ 
    \item In the case where $g$ possesses a (partially) heavier tail than $Y$ in the sense that
    $\limsup_{u \to \infty} \PP(\gu(\bm X) = +1)/\PP(\Yu = +1) = \infty$, we also see by Equation \eqref{eq:ru-rewritten} that  
     $$ \limsup_{u \to \infty} \Ru(g) =  1. $$ 
\end{itemize}}
This indicates that,  whenever $g$ and {\color{black} $Y$} are {\color{black} tail comparable, but} not tail equivalent, the classifier $g$ cannot outperform naive classifiers with respect to the risk $\Ru(g)$ for large $u$.
 While tail equivalence can be related to marginal properties of the classifier and the observations,
the extremal risk also depends on the dependence between those two which is measured in terms of $\chi^*(g)$, as the following lemma shows.

\begin{lemma}\label{lem: R(g)}
 Assume that the two limits $c(g) \in (0,\infty)$ and $\chi^*(g) \in [0,1] $
 defined in Definition~\ref{def: tail equivalent} exist. Then, $\chi^*(g) \leq c(g)$ and the extremal risk has the expression 
\begin{equation} \label{eq:R-xi-c}
  R(g)  = 1 -\frac{\chi^*(g) }{1+c(g)  - \chi^*(g)}. 
\end{equation}
In particular, 
 $
R(g)=0 \mbox{  if and only if }  c(g)=\chi^*(g)=1.
$
\end{lemma}

This lemma indicates that  
$c(g)=1$\footnote{Within the framework of forecast verification, $c(g)=1$ means that $g$ and $Y$ are \textit{asymptotically calibrated}.} is a necessary condition to have $R(g)=0$.
Note that this condition implies that  the constant  $\chi^*(g)$ simply corresponds to the ``classical'' tail dependence coefficient
$\chi(g)$ of the events $\{\gu(\bm X) = +1\}$ and $\{\Yu = +1\}$ as $u \to \infty$, \textit{i.e.}
$$ \chi(g) := \lim_{u \to \infty} \PP(\gu(\bm X)=+1 \mid \Yu=+1) = \lim_{u \to \infty} \PP(\Yu=1 \mid \gu(X)=+1), $$
and so, the case $c(g)=1$ simplifies the expression of the risk  
\begin{equation} \label{eq:R-xi}
R(g)  = 1 - \frac{\chi(g) }{2 - \chi(g)}.
\end{equation}
This equality tells us that any {asymptotically} calibrated classifier with $\chi(g)=0$ always produces a risk function $R(g) = 1$.
Consequently,  any asymptotically independent classifier is as uninformative as the two naive classifiers. 
A reasonable  strategy will be  to dismiss all asymptotically independent classifiers  and find/construct new asymptotically dependent classifiers with positive $\chi(g)$. 
But, finding asymptotically dependent  classifiers  can be complex  in practice, and in addition, in some not so exotic setups, this is not always possible. 
To see this, we  consider the simple non-linear regression  model in the following lemma.

\begin{lemma}\label{lem: R additive}
Assume that the observations $\Yu$ are generated by a non-linear regression  model: for all $u \geq 0$,
\begin{equation*}
 \Yu = \begin{cases}
+1, & \mbox{if } f(\bm X) + N > u\\
-1, & \mbox{otherwise}
\end{cases},
\end{equation*}
where $N$ corresponds to  a {\color{black} regularly varying} random  noise, \textit{i.e.} the function $x\mapsto\PP(N>x)$ is regularly varying at infinity, and $\bm X$ corresponds to the explanatory variables, independent of $N$.
If 
$$
\PP(f(\bm X) > u)= o\left( \PP(N > u) \right) \quad (u \to \infty), 
$$
then for any  level-dependent classifier $g$, we always have  
 $ R(g)  = 1 $
 provided that $R(g)$ exists.
Hence, no classifier can outperform naive classifiers for this regression model. 
\end{lemma}

Note that even if  the forecaster knows exactly the function $f(\cdot)$  and has drawn from the explanatory $\bm X$, the ``ideal'' classifier  
$$ \gu(\bm X) = \begin{cases}
+1, & \mbox{if } f(\bm X) > u\\
-1, & \mbox{otherwise}
\end{cases}, $$
will perform badly, \textit{i.e.}\ $R(g)  = 1$. 
In addition, the classical trick  of using ranks to avoid the problem of marginals discrepancy cannot be applied here. 
For example, suppose that  $f(\bm X) + N$ is unit Fr\'echet distributed, then transforming 
the marginals of $\bm X$ into unit Fr\'echet random variables, say into $\bm \tilde{\bm X}$, does not 
remove the issue as    the unobserved noise  $N$  still has heavier tails than  $\tilde{f}(\bm{\tilde{X}})=f(\bm X)$ for some function $\tilde{f}(\cdot)$.  An example of a simulation of this framework is given in Section \ref{sec::simulations} (see also Figure \ref{fig::simulation}).
So, a finer risk measure is needed to distinguish different classifiers in case of asymptotic independence, as defined in Definition \ref{def: tail equivalent}, between the classifiers and the observations $Y$.

\subsection{The extremal conditional risk $\Reps$}
\label{subsec:extremalconditionalrisk}

The choice of the conditions in
Lemma \ref{lemma: sets} brings new possibilities to construct  finer risk measures for extremes than $\Ru$.
 In particular, the conditions allow us to exclude events that are not at least ``a little extreme'' (w.r.t. some smaller level $v$) both in terms of the observation $\Yv$ and the classifier $\gv$.
Such a modeling strategy  is at the core of hidden regular variation and asymptotic independence models. 
More precisely,   we first need  to fix  marginal features. 
We assume 
that both   $u \mapsto \PP(\gu(\bm X)=+1)$ and $u \mapsto \PP(\Yu=+1)$ are regularly varying with indices $\alpha_g>0$ and $\alpha_Y>0$, respectively. 
This means that for any  $\varepsilon \in (0,1)$,  
$$ \lim_{u \to \infty} \PP(\gu(\bm X) = +1 \mid g^{(\varepsilon u)}(\bm X) = +1) = \varepsilon^{\alpha_g} 
\mbox{ and }
 \lim_{u \to \infty} \PP(\Yu = +1 \mid Y^{(\varepsilon u)} = +1) = \varepsilon^{\alpha_H}.$$
 These limits have to be understood in terms of Equation \eqref{eq: mixing cons}, \textit{i.e.}\ in terms of
  $ \PP( \gu(\bm X) = +1 \mid \gv(\bm X) = +1)$ and $\PP( \Yu=+1 \mid \Yv=+1)$ with $v=\varepsilon u$.
 To apply  \eqref{eq: mixing cons}, the mixing condition \eqref{eq: mixing} needs to be satisfied. 
To do so,  we opt for an extended 
version of the framework of \cite{Ramos2009}, \textit{i.e.}\
\begin{equation}\label{eq: ramos}
  \PP[\gu(\bm X) = +1, \Yv =  +1 ] = L(u,v) (u^{-\alpha_g} v^{-\alpha_Y})^{1/2\eta},
\end{equation}
where $\eta \in (0,1]$ indicates the rate of decay of the joint probability as a function of $u$ and $v$ and $L(\cdot,\cdot)$ is bivariate slowly varying function, \textit{i.e.}\ there exists a limit function $\ell^*: (0,\infty) \times (0,\infty) \to (0,\infty)$ defined as
$$ \ell^*(s,t) = \lim_{u \to \infty} \frac{L(us,ut)}{L(u,u)}, \qquad s,t > 0 $$
and satisfying $\ell^*(cs, ct) = \ell^*(s,t)$ for all $c, s, t > 0$.
The parameter $\eta$ measures the  extremal dependence strength.
The case $\eta=1$ (and $\lim_{u \to \infty} L(u,u) > 0$) corresponds to the asymptotic dependence of $Y$ and $g$ while $\eta<1$  (or $\eta=1$ and $\lim_{u \to \infty} L(u,u) = 0$) corresponds to the asymptotic independence case. 
In particular, if $\eta=0.5$, then  independence appears in the extremes. 
If $0.5 <\eta<1$ ($0 <\eta <0.5)$ the extremes are said to be  positively (negatively) associated.
 
Now, noticing that the joint probability in Equation \eqref{eq: ramos} corresponds to the mixing condition \eqref{eq: mixing}, we can apply Lemma \ref{lemma: sets}, see Appendix \ref{sec: proofs} for a proof. 

\begin{proposition}\label{prop: R eps}
Under the Ramos and Ledford model defined  by \eqref{eq: ramos}, the extremal conditional risk in \eqref{eq:reps} can be expressed as 
$$ R_\varepsilon(g) = 1 - \frac{1}{{\ell^*(\varepsilon,1)} \varepsilon^{- \alpha_g/2\eta} + {\ell^*(1,\varepsilon)} \varepsilon^{- \alpha_Y/2\eta}  - 1} \quad
\mbox{ for any } \varepsilon \in \mo{(}0, 1)
$$
 such that $\ell^*$ is continuous at $(\varepsilon,1)$ and $(1,\varepsilon)$.
\end{proposition}

For fixed $\varepsilon \in \mo{(}0,1)$, the risk function $\Rbar_\varepsilon(g)$ decreases with increasing $\eta$. So, given all parameters are fixed but $\eta$, the forecaster should aim  at maximizing $\eta$.
In practice, two forecasters, say $g_1$ and $g_2$, may produce different $\ell^*(\cdot,\cdot)$ and $\alpha_g$.
Consequently, the minimization  of $\Rbar_\varepsilon(g)$  can also depend,  besides $\eta$, on other parameters. 

\begin{table}[h]
\caption[]{\label{table::risks} Overview of the risk measures used in this study.}
\vspace{3mm}
\centering
\begin{tabular}{cccl}
\toprule
Notation  & Usage & \multicolumn{2}{c}{Associated limiting risk} \\
\midrule
$\Ru(g)$ \eqref{eq:Ru} & Asymp. dependence & $R(g) =  \lim\limits_{u \to \infty} \Ru(g)$ 
& \textit{Extremal risk} \\
\midrule
$\Ruv(g)$  \eqref{eq:Ruv} & Asymp. independence & $\Reps(g) = \lim\limits_{\substack{u,v \to \infty\\ v/u \to \varepsilon}} \Ruv(g)$ & \textit{Extremal conditional risk} \\
\bottomrule
\end{tabular}
\end{table}

\subsection{Risk function inference}\label{sec: emp risk}

Concerning the estimation of $\Rbar(g)$ and $\Rbar_\varepsilon(g) $ defined by \eqref{eq:reps}, the empirical estimator can be easily computed from the sample 
$(\gu(\bm X_i), \Yu_i, \gv(\bm X_i), \Yv_i)_{i=1,\dots,n}$
based on two thresholds $u,v$ such that $v/u \approx \varepsilon$. 
More precisely, such an estimator is given by
\begin{equation}\label{eq: emp class}
    \Rhatuv(g) = \frac{\sum_{i=1}^n \one\{\gu(\bm X_i) \neq \Yu_i, \Yv=1, \gv(\bm X_i) = 1\}}{\sum_{i=1}^n \one\{\max\{\gu(\bm X_i), \Yu_i\}=1, \Yv_i=1, \gv(\bm X_i) = 1\}}.
\end{equation}
The following proposition describes the asymptotic property of such an estimator. \mo{Note that it holds also for the case $\varepsilon=0$ where one  considers the sequence $v_n \equiv 0$.}

\begin{proposition} \label{prop:emp_est} 
\mo{Let $\varepsilon \in [0,1)$ and} assume that the risk function $\Rbar_\varepsilon(g)$ defined by \eqref{eq:reps} exists.
Consider \mo{a sequence $u_n \to \infty$ and a sequence $v_n \equiv 0$ (if $\varepsilon=0)$ or $v_n \to \infty$  such that $v_n / u_n \to \varepsilon > 0$.  Furthermore, we assume that $n p_{g}(u_n,v_n) \to \infty$ as $n \to \infty$} where
$$
p_{g}(u_n,v_n) \coloneqq \PP(\max\{\gun(\bm X), \Yun\}=1, \gvn(\bm X) = \Yvn = 1).
$$
If 
$$ \lim_{n \to \infty} \sqrt{n p_{g}(u_n,v_n)} \left( \frac{\PP(\gun(\bm X) \neq \Yun,
\gvn(\bm X) = \Yvn = 1)}{p_{g}(u_n,v_n)} - \Rbar_\varepsilon(g) \right) = 0, $$
then the empirical estimator $\Rhatunvn(g)$ {\color{black} is asymptotically normal with negligible bias, \textit{i.e.},}
$$ \sqrt{n p_g(u_n,v_n)} \left( \Rhatunvn(g) - \Rbar_\varepsilon(g) \right) \xrightarrow[n\to\infty]{d}
\mathcal{N} \left(0, \Rbar_\varepsilon(g) (1-\Rbar_\varepsilon(g)) \right). $$
\end{proposition}

\begin{remark}
{\color{black} From the proof of Proposition \ref{prop:emp_est} it follows that 
$$ \frac{ \frac 1 n \sum_{i=1}^n \one\{\max\{\gu(\bm X_i), \Yu_i\}=1, \Yv_i=1, \gv(\bm X_i) = 1\}}{p_g(u_n,v_n)} \longrightarrow 1 $$
in probability as $n \to \infty$, i.e., the numerator is a consistent estimator of $p_g(u_n,v_n)$.
As $p_g(u_n,v_n)$ occurs in the rate of convergence in Proposition \ref{prop:emp_est}, this estimator can be used to assess the variance of $\Rhatuv(g)$.
}\end{remark}

{\color{black} While Proposition \ref{prop:emp_est} provides asymptotic normality of an estimator for the extremal (conditional) risk \mo{$\Reps(g)$} for a single classifier $g$, the comparison of the extremal conditional risks of two classifiers $g_1$ and $g_2$ requires joint asymptotic normality of the estimators $\Rhatunvn(g_1)$ and $\Rhatunvn(g_2)$. In order to obtain such a result, some more assumptions on the joint behavior of
$\gu_1(\bm X)$ and $\gu_2(\bm X)$ are needed.  A statistical test for a comparison that has been developed, can be found in Appendix \ref{appendix::test}. It has been used in both our simulation study and the application study.

\section{Simulations}
\label{sec::simulations}

\subsection{A simple linear setup}\label{sec: linear}

 In the following, the necessary ingredients as in Table \ref{table::ingredients} are computed  by truncating continuous responses. This approach is aimed at obtaining the necessary input variables and should not be considered as the reference framework for the application of the two  developed extremal risk measures.

Our main simulated example is  therefore based on a simple linear regression model but with the feature that the components of the input vector  $\bm {X}$ do not have the same tail behavior and the noise is regularly varying, see Lemma 
\ref{lem: R additive}. 
 More precisely, the multivariate vector $\bm{X} = (X_1, X_2, X_3, X_4)$ consists of independent components where:
\begin{equation*}
X_1 \sim \mathrm{Pareto}(3),\mbox{ }  X_2 \sim \mathrm{Pareto}(2), \mbox{ } X_3 \sim \mathrm{Exp}(1),\mbox{ } X_4 \sim \mathrm{Exp}(2)
\end{equation*}
Here, $X_1$ and $X_2$ follow Pareto distributions with tail indexes 3 and 2 respectively, while $X_3$ and $X_4$ follow exponential distributions with scale parameters 1 and 2 respectively.
 For the simulations, we use the same framework as in Lemma \ref{lem: R additive}, \textit{i.e.} assuming that the events of interest are given by
\begin{equation}\label{eq::simulation} \Yu \coloneqq \left\{\begin{aligned}
+1 & \mbox{, if } X_1 + N > u\\
-1 & \mbox{, otherwise}
\end{aligned} \right.,\end{equation}

 for $u \geq 0$, where $N\sim \mathrm{Pareto}(2)$ represents  an independent noise with heavier tail than $X_1$. 
So, given a sample $(\{X_{j,i}\}_{1\leq j\leq 4},N_i)_{i=1,\dots,n}$ with $n=10000$, our goal is to compare different classifiers in terms of predicting extreme occurrences, here defined as the event
 $\{ X_1+N>u \}$ with $u$ equal to the  $97$th percentile of  $X_1+N$. In this simulation setup, it is clear from \eqref{eq::simulation} that all  variables but  $X_1$ are useless to  explain the extreme occurrences of $\Yu$. 
In addition, Lemma 
\ref{lem: R additive} tells us that the relevant information contained in the  variable $X_1$ is hidden by the heavier noise $N$, \textit{i.e.}\ we are in the case of asymptotic independence. 

An example of such simulation is given in Figure \ref{fig::simulation}. 
The left panel displays a scatter plot between the response variable $f(\bm X)+N$ (left axis) and the explanatory variable $f(\bm X)$ (right axis).  In the present simulation framework, $f(\bm X)=X_1$. As expected, 
no sign of asymptotic dependence can be found in the upper corner. In the right panel, 
we remove the mass along  the axis (gray points)  by conditioning  on the joint event  $\{X_1 > v\} \cap \{\Yv=+1\}$ with $v=\varepsilon u$ and $\varepsilon=0.7$, \textit{i.e.} we consider the ``ideal'' classifier $\{\gv(\bm X) = +1\} \coloneqq 
\{X_1 > v\}$. 
The right panel zooms on these specific points and highlights  a clear  dependence between $X_1+N$ and $X_1$ that was hidden by the heavy tailed noise $N$ in $X_1 + N$.

\begin{figure}[h]
   \begin{center}
    \includegraphics[scale=0.4]{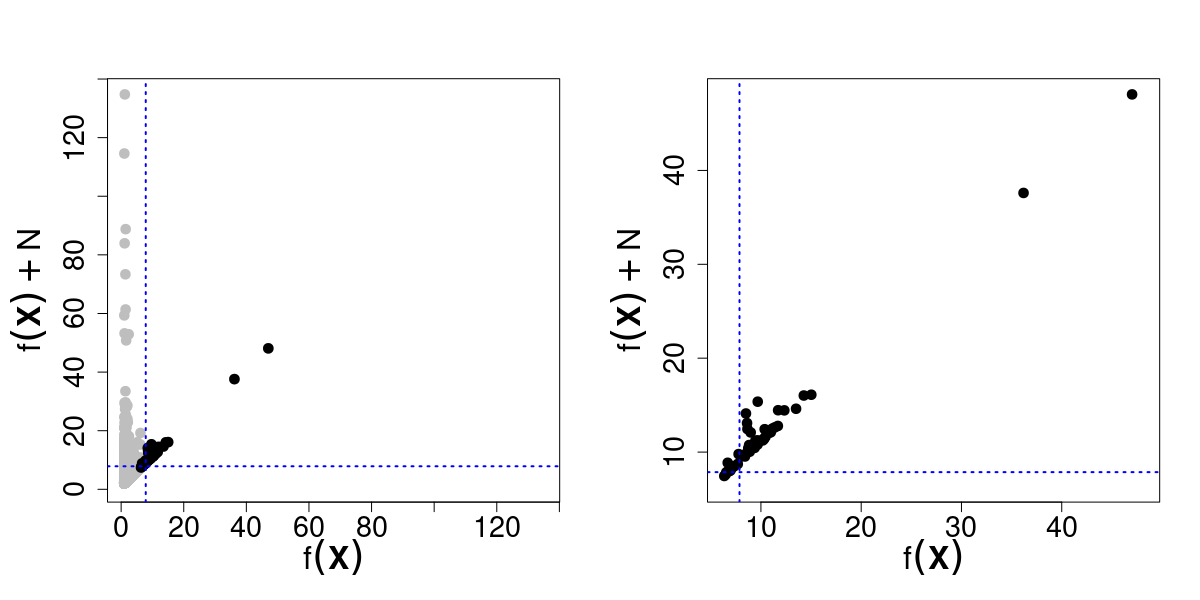}
    \end{center}
    \caption[Simulated example for Lemma \ref{lem: R additive}]{Simulated example of the framework defined in Equation \eqref{eq::simulation}. This figure is also an illustration for Lemma \ref{lem: R additive}.}
    \label{fig::simulation}
\end{figure}

In practice, we do not know  the ideal choice of classifier $g$, so we  need to introduce different classifiers and compare them.

\subsection{Classifiers descriptions}
\label{sec::classif_descr}
Table \ref{table::methods} (see Appendix \ref{sec:classif}) provides the list of classifiers that we compare with our metrics \eqref{eq: emp class}. 
This list contains some of the  most standard classifiers found in the literature \cite[see, e.g.,][]{Hastie2009}:
logistic regression (\textit{Lasso}), decision tree (\textit{Tree}), random forest (\textit{RF}) and support vector machine (\textit{SVM}). We apply them  with their built-in cost function that is not necessarily fine-tuned to forecast   extremes. This is not an issue because our main goal is  to compare existing forecasters, and not to create new ones \cite[see, e.g.,][for such developments]{Jalalzai2018, Wang2010}. An example of binary outputs from the decision tree classifier is given in Table \ref{table::counts_tree}.  The difference between the left and right panels corresponds to the training set based on  two different thresholds $u$ and $v$, \textit{i.e.} the ingredients needed in our study. 

\begin{table}[h]
\caption{\label{table::counts_tree} Contingency table depicting an example of predicted binary output for the decision tree classifier with two different thresholds $v$ and $u>v$, with $v=0.4u$. The results shown are computed upon a testing set whose sample size is equal to 3000 ($30\%$ of the data)}
\vspace{3mm}
\centering
\begin{tabular}{ccc}
\toprule
 Threshold $v$ & $Y^{(v)}=+1$ & $Y^{(v)}=-1$ \\
 \midrule
 $g^{(v)}(\bm X) = +1$ & $646$ & $213$ \\
 $g^{(v)}(\bm X) = -1$ & $662$ & $1479$ \\
 \bottomrule
\end{tabular}
\hspace{1.5em} 
\begin{tabular}{ccc}
\toprule
 Threshold $u>v$& $Y^{(u)}=+1$ & $Y^{(u)}=-1$ \\
 \midrule
 $g^{(u)}(\bm X) = +1$ & $12$ & $0$ \\
 $g^{(u)}(\bm X) = -1$ & $66$ & $2922$ \\
 \bottomrule
\end{tabular}
\end{table}

 In order to fix a baseline in terms of performance, we compare these four classic classifiers with a linear classifier
 defined as follows
$$ \gu_{\bm \theta}(\bm X) = \begin{cases}
  +1, & \quad \bm \theta^\top \bm X > u, \\
 -1, & \quad \bm \theta^\top \bm X \leq  u,
\end{cases} \qquad  \text{for some } \bm \theta \in [0,\infty)^d.  $$
{\color{black} This specific classifier is discussed in more details in Appendix \ref{sec:lin_class}. In short, Proposition \ref{proof: RV and unique mimizer} shows that under mild conditions, one can find some optimal $\bm\theta^\ast$ such that the associated extremal risk $R(g_{\bm\theta^\ast})$ is minimal. In such context, and given the simulation framework \eqref{eq::simulation}, we expect the optimal linear classifier $\gu_{\bm\theta^\ast}$ to be the best among the different classifiers considered.}

\subsection{Implementation and results} 
We split our simulated  data set in two: $70\%$ for a training part, over which we train our different classifiers  to get good predictive power; $30\%$ for a testing part, which we use to estimate {\color{black} different} risks $R^{(u,v)}$.  Since we consider here simulated data, we assume that we can artificially generate several data sets as in Table \ref{table::ingredients} for a range of thresholds $v$. 
Note that each algorithm has the same inputs, in particular the  same binary sequence describing  the events  $\{Y^{(u)}=+1\}$
with $u$ set to be equal to the $97$th percentile of  $X_1+N$,  and $\{Y^{(v)}=+1\}$ with $v=\varepsilon u$ for different values of $\varepsilon$.
This cross-validation procedure has been repeated $50$ times.
The sample used to compute our risk functions is based on the bivariate binary vectors  $(g^{(u)}(\{X_{j,i}\}_{1\leq j\leq 4}),\{Y^{(u)}_i =1\})_{i=1,\dots,n}$.
In addition, the binary outputs of the classifiers $g$ are obtained under the threshold $u$ and some thresholds $v<u$, so the training part has to be performed twice (once for each threshold). 
Then,   the empirical risk estimator defined  by \eqref{eq: emp class} can be computed. 
Figure \ref{fig::simuRisk}  
shows the sensitivity of the classifier ranking with respect to the value of $v$.

\begin{figure}
\centering
\includegraphics[scale=0.4]{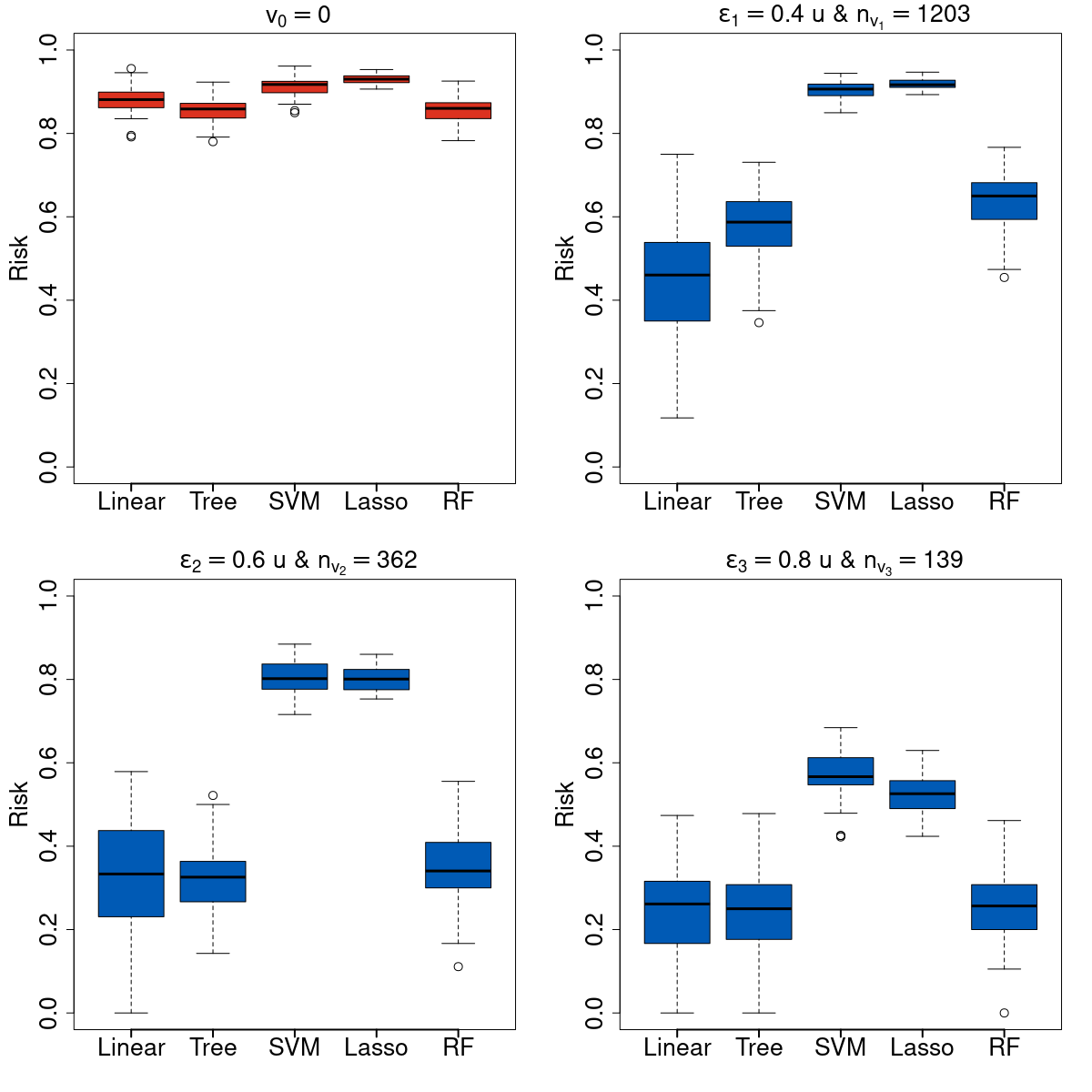}
\caption[Estimation of $R_\varepsilon(g)$ for different classifiers]{Estimation of $R^{(u,v_k)}(g)$ for different classifiers $g$ (cross-validation with $50$ repetitions) and $k\in\{0,1,2,3\}$. In red (top-left) are the estimates for $v_0=0$ and in blue for different values of $v_k=\varepsilon_k u$, $k>0$ and $\varepsilon_k\in\{0.4, 0.6, 0.8\}$. At the top of each plot, the value of $\varepsilon_k$ and the number of points $n_{v_k}$ such that $\{Y^{(v_k)}=+1\}$ from the testing set are given.}
\label{fig::simuRisk}
\end{figure}
 According to Lemma \ref{lem: R additive}, our five classifiers cannot asymptotically outperform the naive classifiers. And as $u$ gets to infinity, the estimated risks should converge to $1$, the worst possible value. The top-left panel in Figure \ref{fig::simuRisk}, corresponding to the case $v = 0$, shows that each classifier has a risk near to one at high, but finite, level $u$. So, to better discriminate classifiers, we
need to remove the masses along the axes by setting a positive value for $v$.
As we increase $v$, the size of sets needed to compute   \eqref{eq: emp class} becomes smaller, see the values of $n_{v_k}$ in the legend of each panel.
Hence, the blue box plots become wider as $v$ increases:   a classical bias-variance trade-off.  
As we know from \eqref{eq::simulation} that the true generative  process is linear, 
$v=0.4u$ appears as a reasonable value to balance the bias-variance trade-off. In this case, our  linear classifier, tailored to handle  linear asymptotic independence cases, outperforms the other classifiers. And among these four other classifiers, the decision tree appears to be the best, but it is still far from the optimal linear solution.
More importantly, for this specific empirical study the overall ranking seems to be insensitive to the values of  $v>0$.
Other simulations concerning the regular variation case are available upon request.

\section{Danube river discharges}\label{sec: Danube river discharges}

We now apply our assessment approach to  summer daily   river discharges (measured in $m^3/s$) at $31$ stations spread over the upper Danube basin (see Figure \ref{fig::Danube} in Appendix \ref{sec::appRiverDep}), and recorded over the time period 1960-2010 in June, July and August. 
These observations  have been  studied by the EVT community  \cite[see, e.g.,][]{Asadi2015,Mhalla2020,Gnecco2021}.
This dataset was made available by the Bavarian Environmental Agency (\url{http://www.gkd.bayern.de}).

To remove temporal clustering in extreme river discharges, 
\cite{Mhalla2020} in their Section 5 implemented a declustering step,  we use here these declustered data. 
Each station then  contains $n=428$ observations   that we will consider  temporally independent. 
In order to  reduce the large discrepancy in terms of discharges magnitude among stations, we force  the starting value of all $31$ time series to  equal  zero by subtracting  to  each station its minimum. 
Then, we re-normalize each time series by its range  (\textit{i.e.}\ the difference between the maximum and the minimum of each time series).

These post processing treatments  are useful to display and interpret  the data at hand and do not impact the classifiers performance.

Although all $31$ station recordings are available, we can artificially remove one station and try to predict its extreme occurrences from a given subset of other weather stations. 
In this section, we remove station $1$ (downstream) and try to predict its  extreme occurrences from only stations $23$ and $24$, which are indirect tributaries to the main 
river flow. 
So, this setup is complex\footnote{Section \ref{sec::appRiverDep} treats a simpler case where the extreme occurrences at station $1$ are predicted from the whole set of remaining stations. 
In this case, strong dependencies among station $1$ and other stations  can be observed. So, the main issue is to select these stations, a problem discussed in Section \ref{sec:lin_class}.} 
for two reasons. First,  station $1$, as a downstream point that accumulates  all discharges, has a much heavier tail than the two  tributaries. 
Second,  empirical investigations (not reported here) show that it is difficult to determine if we are in the asymptotically dependent or independent case.

 As in Table \ref{table::ingredients}, we need to have at our disposal the occurrences of extreme river discharges at station 1 for two different thresholds $v$ and $u$ (with $v<u$). Note that we could assume that we do not observe the whole time series of river discharges at this station, only the extreme occurrences for two different levels. Furthermore, we could also assume that we only have outputs from different classifiers, as in Table \ref{table::ingredients}, previously built using the observations of stations 23 and 24. However, in the following, we assume that we can observe the whole time series of river discharges at stations 23 and 24, that we use to built  different classifiers and then compare their performances.

Figure \ref{fig::hrv_river2_res} 
summarizes our findings.
 Considering a conditioning set with the lower threshold $v$ implies that only $190$ points remain from the original length of $428$ data points per   station.
This can explain why, looking at Figure \ref{fig::hrv_river2_res}, the uncertainty in the risk estimate increases when considering $R^{(u,v)}$ (blue boxes, right panel) instead of  $R^{(u,0)}=R^{(u)}$ (red boxes, left panel).

\begin{figure}
\centering
\includegraphics[scale=0.35]{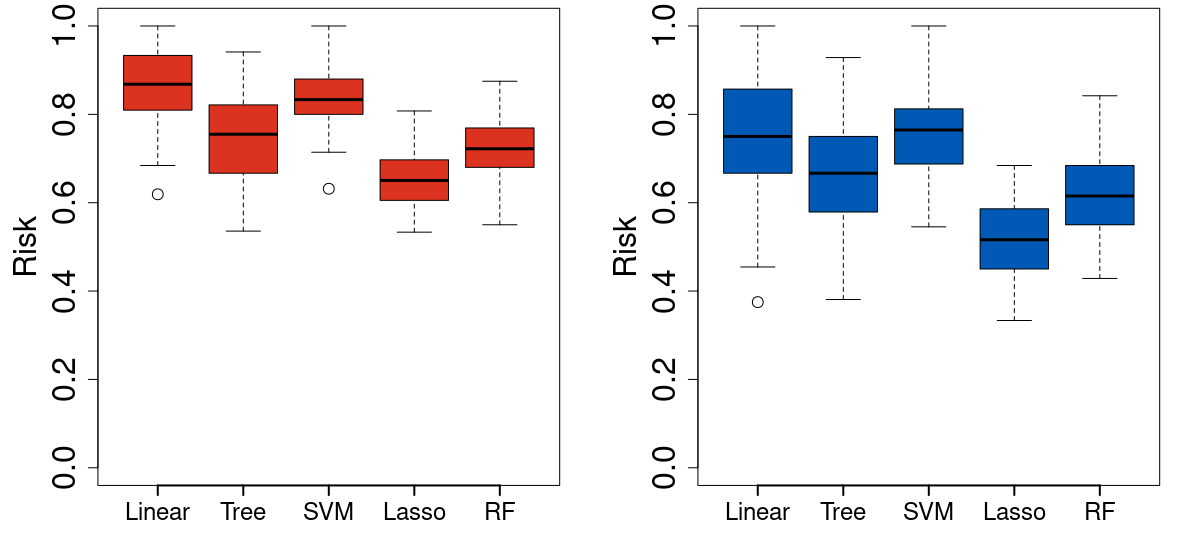}
\caption[Estimation of $R_\varepsilon(g)$ for five different classifiers with the Danube river discharges]{Estimation of $R^{(u,v)}(g)$ for five different classifiers (cross-validation with $50$ repetitions, $70\%$ train and $30\%$ test), threshold $u$ is the $0.85$ quantile of $X_1$. In red (left) are the estimates when $v=0$ and in blue (right) for $v=0.6 u<u$. 
The length of the testing set is equal to $129$, this leads to around $60$ points such that $\{Y^{(v)}=+1\}$ and nearly $20$ points such that $\{Y^{(u)}=+1\}$.}
\label{fig::hrv_river2_res}
\end{figure}

 First, unlike the  simulation example in Section \ref{sec: linear}, we do not want to assume here that we can observe extreme occurrences at station 1 at any level $v$. Furthermore, and as already mentioned, it is not clear to assess  whether our river discharges analysis using the  three selected  weather stations belongs to   the framework of asymptotic independence or not.  
Still, it is reassuring that the ranking of the classifiers in Figure \ref{fig::hrv_river2_res} appears to be insensitive   to the values of   $R^{(u)}$ or $R^{(u,v)}$,  
\textit{i.e.} whether the data are asymptotically dependent or not. 
This hints  that, among all the classifiers, the logistic regression with lasso penalty seems to perform better  than the four other classification methods. 
 This ranking of classifiers is specific to this particular example. No general conclusions about lasso techniques for extremes should be drawn. 

Besides this river example,  we advocate practitioners to compute  risk functions that can both handle  the asymptotic dependence and  asymptotic independence cases.
 This also complements the recent tools used to discriminate between the two cases 
\cite[see, e.g.,][]{Ribereau2022}.
In addition, the linear classifier could provide a simple benchmark with well understood properties with respect to $R^{(u)}$, see Proposition \ref{proof: RV and unique mimizer}.

\section{Discussion}

 This work focused on the elaboration of risk measures tailored for binary classification of extreme events. The main hypothesis of this study is that the ingredients we are working with are in the form of Table \ref{table::ingredients}, \textit{i.e.} we have at our disposal binary outputs of extreme events based on different thresholds. 
 In case of datasets for which  binary classifiers responses can be easily obtained from different $u$ and $\varepsilon$,  we strongly advice to  compute our risk function  $\Rhatuv(g)$ for different thresholds.  Still, caution must be exercised in comparing such risk results. 
As in any extreme value analysis, the choice of  thresholds is delicate. Overall, we can say if one classifier appears,  over a wide range of high thresholds,  to constantly outperform  its competitors  with respect to the metrics we propose, then such a classifier should be favored.  
In contrast, if classifier rankings strongly differs with threshold variations, then caution should be applied. This may mean that all classifiers are equally good or bad. 
An interesting avenue of research will be to determine how to combine classifiers like how to perform Bayesian modeling averaging with respect to our metrics and threshold variations \cite[see, e.g.,][]{Sabourin13}.

Finally, we would like to point out that even if we assume that the thresholds are imposed by a given application, the subject of optimizing an $\varepsilon$ could be addressed in a future work.

\section{Supplementary Materials}
\label{sec::marco fin}
A R package is available on GitHub that implements the empirical estimation of the risk function developed in this paper %(\url{https://github.com/jlegrand35/ExtremesBinaryClassifier}) 
and can be used either to reproduce the results of the conducted classifier comparisons or to perform new comparisons using other binary classifiers.
The data used in the application are available in the R package \texttt{graphicalExtremes} \citep{graphicalExtr}.

% \section*{Acknowledgment}

% Part of this  work was supported by the  DAMOCLES-COST-ACTION on compound events, the French national program (FRAISE-LEFE/INSU
%  and  80 PRIME CNRS-INSU), and the European H2020 XAIDA (Grant agreement ID: 101003469) and funded by Deutsche Forschungsgemeinschaft (DFG, German Research Foundation) under Germany's Excellence Strategy - EXC 2075 - 390740016. 
%  The authors also acknowledge the support of the French Agence Nationale de la Recherche (ANR) under reference ANR-20-CE40-0025-01 (T-REX project), 
%  the ANR-Melody and the Stuttgart Center for Simulation Science (SimTech).

 \bibliographystyle{apalike}
\bibliography{bibliography.bib}
\newpage

\appendix
\section{\textcolor{black}{Overview of different risk measures for imbalanced binary data}}\label{sec:review_risks}

\begin{table}[!h]
\caption[Key features of some known classifiers]{\label{table::review_risks}Key features of some risk measures for extremes. 
The integer $n$ denotes the sample size, $h$ the number of ``hits" (true positive), $f$ the number of ``false alarm", $m$ the number of misses (false negative) and $r$ the number of correct rejection (see Table \ref{table::counts}).
The third column  refers to the type of dependence  when $u$ is large, see Definition \ref{def: tail equivalent}.
The last column focuses on the asymptotic properties of the proposed risk measure estimator.}
\vspace{3mm}
\centering
% landscape \begin{tabular}{p{2.5cm} p{6cm} p{7cm} l}
\begin{tabular}{p{3cm} | p{4.5cm} |  p{4.2cm} | p{3cm}l}
\toprule
Name & Definition & Asymptotic extremal dependence  type & Estimators  Properties\\
\midrule
Critical success index \citep{Schaefer1990} &  $\dfrac{h}{h+f+m}$ &  Dependent case & To be covered in this work\\
\midrule
Asymptotic risk in the extremes  & 
$\PP(f+m \mid \|\bm X\| > u)$ where $\|\bm X\| > u$ defined extremes in the covariate space $\bm X$
% Classifiers % $g(\bm X)$ 
% minimizing the limit  
%Consider extremes as rare events from the binary classifier outputs;
&  Dependence case & Derived in \citep{Jalalzai2018} \\
% & 
% %$\PP(Y\neq g(\bm X) \mid \|\bm X\| > u)$ 
% %$f+m \mid \|\bm X\| > u$
 
% &   &\\
\midrule
Extreme dependency score  & $\dfrac{2\log\left[(h+m)/n\right]}{\log(h/n)}-1$ & Independence case & NA in \citep{Stephenson2008}\\
\midrule
Extremal dependence index  & $\dfrac{\log\frac{f}{f+r}-\log\frac{h}{h+m}}{\log\frac{f}{f+r}+\log\frac{h}{h+m}}$ & Independence case &   NA in \citep{Ferro2011} \\
\midrule
Extension of the critical success index &  see \eqref{eq:Ruv} &  Independent case & To be covered in this work\\
%\midrule
%Adjusted F-score \citep{MARATEA2014}& &  &\\
\bottomrule
\end{tabular}
\end{table}
%\end{landscape}

\begin{table}[h!]
\caption{\label{table::counts} Contingency table representing the different outputs for binary classification.}
\vspace{3mm}
\centering
\begin{tabular}{ccc}
\toprule
 & $Y^{(u)}=+1$ & $Y^{(u)}=-1$ \\
 \midrule
 $g^{(u)}(\bm X) = +1$ & $h$ & $f$ \\
 $g^{(u)}(\bm X) = -1$ & $m$ & $r$ \\
 \bottomrule
\end{tabular}
\end{table}

\newpage
\section{Overview of the different classifiers considered in Section \ref{sec::simulations}}\label{sec:classif}

\begin{table}[!h]
\caption[Summary and key features of different classifiers]{\label{table::methods}Summary and key features of the different classifiers studied. See for example  \cite{Hastie2009} for a comprehensive review of the last four classification methods.}
\centering
\begin{tabular}{l p{10cm}}
\toprule
Method & Main features \\
\midrule
Linear classifier & Simple binary classifier, parameters estimation based on  the minimization of the risk function over the set of contributing variables, theoretical value of $R(g_\theta)$ inferred  from spectral decomposition.\\
\midrule
Logistic regression & Parametric linear model with a lasso penalty,  coefficients of less contributing variables are set to zero. \\
\midrule
Decision trees & Easy to interpret, gives relative importance of each variables, learns simple decision rules inferred from the input. \\
\midrule
Random forests &  Builds multiple decision trees combined by majority vote,  better predictive power than decision trees. \\
\midrule
Support vector machines & Finds the best hyperplane to separate two overlapping classes, generally performs  better than the other classifiers. \\
\bottomrule
\end{tabular}
\end{table}

\newpage
\section{Testing the equality of risk metrics for two classifiers}\label{appendix::test}

\mo{In the following, we want to generalize Proposition \ref{prop:emp_est} in order to obtain the joint asymptotic behavior of the empirical risks of two extremal classifiers $g_1$ and $g_2$.
Similarly to Proposition~\ref{prop:emp_est}, the result covers the case $\varepsilon=0$ if we choose $v_n \equiv 0$.}

\mo{Furthermore, we note} that the rates $(n p_{g_1}(u_n,v_n))^{1/2}$ and $(n p_{g_2}(u_n,v_n))^{1/2}$ might be different. Thus, joint convergence holds at a potentially slower rate
$[n (p_{g_1} \wedge p_{g_2})(u_n,v_n)]^{1/2}$. In this case, it might happen that one of the asymptotic variances is equal to zero.

\begin{proposition}\label{prop:emp_est_biv}
Let $g_1$ and $g_2$ be two extremal classifiers and $\varepsilon \in [0,1)$.
Assume that the both risk functions $\Rbar_\varepsilon(g_1)$ and $\Rbar_\varepsilon(g_2)$ exist as well as the two limits
$$ \displaystyle \cepsone \coloneqq \limeps \frac{p_{g_1}(u,v) \wedge p_{g_2}(u,v)}{p_{g_1}(u,v)} \in [0,1] \quad \text{and} \quad \displaystyle \cepstwo \coloneqq \limeps \frac{p_{g_1}(u,v) \wedge p_{g_2}(u,v)}{p_{g_2}(u,v)} \in [0,1].$$
Furthermore, we assume that the limits $\rreps \in [0,1]$, $\qepsonetwo \in [0,1]$, $\qepstwoone \in [0,1]$ and $\peps \in [0,1]$ given in appendix \ref{sec:appendixF5} exist.

Consider sequences  \mo{$u_n$ and $v_n$} such that $v_n / u_n \to \varepsilon$, \mo{$u_n \to \infty$} and $n (p_{g_1} \wedge p_{g_2})(u_n,v_n)  \to \infty$ as $n \to \infty$.
If 
$$ \lim_{n \to \infty} \sqrt{n (p_{g_1} \wedge p_{g_2})(u_n,v_n)}
 \left( \frac{\PP(\gun_1(\bm X) \neq \Yun, \gvn_1(\bm X) = \Yvn = 1)}{p_{g_1}(u_n,v_n)} - \Rbar_\varepsilon(g_1) \right) = 0 $$
and
$$ \lim_{n \to \infty} \sqrt{n (p_{g_1} \wedge p_{g_2})(u_n,v_n)}
\left( \frac{\PP(\gun_2(\bm X) \neq \Yun,  \gvn_2(\bm X) = \Yvn = 1)}{p_{g_2}(u_n,v_n)} - \Rbar_\varepsilon(g_2) \right) = 0 $$
then the two empirical estimators $\Rhatunvn(g_1)$ and $\Rhatunvn(g_2)$ are jointly asymptotically normal with negligible bias, \textit{i.e.},
\begin{align*}
 \sqrt{n (p_{g_1} \wedge p_{g_2})(u_n,v_n)} &
\left( \begin{array}{c} \Rhatunvn(g_1) - \Rbar_\varepsilon(g_1)\\
                        \Rhatunvn(g_2) - \Rbar_\varepsilon(g_2) \end{array} \right)  \\
                        %\stackrel{n \to \infty}{\longrightarrow}_d 
\xrightarrow[n\to\infty]{d} &
\mathcal{N} \left( \left[ \begin{array}{c} 0 \\ 0 \end{array} \right],
                   \left[ \begin{array}{cc} 
                       \cepsone \Rbar_\varepsilon(g_1) (1-\Rbar_\varepsilon(g_1)) & 
                      \sigma_{12}
                       \\
                       \sigma_{12} & \cepstwo \Rbar_\varepsilon(g_2) (1-\Rbar_\varepsilon(g_2))
                       \end{array} \right]
                       \right),
\end{align*}
where
$$ \sigma_{12} =  \sqrt{\cepsone \cepstwo} \left( \rreps - \qepstwoone \Reps(g_1) - \qepsonetwo \Reps(g_2) + \peps \Reps(g_1) \Reps(g_2) \right) $$
\end{proposition}}

Note that, under the assumption that $\Reps(g_1)=\Reps(g_2)$, 
Proposition \ref{prop:emp_est_biv} implies that 
$$ \sqrt{n (p_{g_1} \wedge p_{g_2})(u_n,v_n)} (\Rhatunvn(g_1) - \Rhatunvn(g_s)) \to_d \mathcal{N}(0, \sigma^2)$$
where
$$ \sigma^2 = \cepsone \Rbar_\varepsilon(g_1) (1-\Rbar_\varepsilon(g_1)) - 2 \sigma_{12} + \cepstwo \Rbar_\varepsilon(g_2) (1-\Rbar_\varepsilon(g_2)).$$
Thus, if $\sigma^2$ is known, approximate $Z$-tests can be used 
to test hypotheses such as $\Reps(g_1) \geq \Reps(g_2)$, $\Reps(g_1) =\Reps(g_2)$ or $\Reps(g_1) \leq \Reps(g_2)$. In practice, $\sigma^2$ needs to be estimated, which can be done by plugging the empirical counterparts of $\Reps(g_1)$, $\Reps(g_2)$, $\cepsone$, $\cepstwo$, $\peps$, $\rreps$, $\qepsonetwo$ and $\qepstwoone$ into the formula for $\sigma^2$.

\begin{figure}
\centering
\includegraphics[scale=0.4]{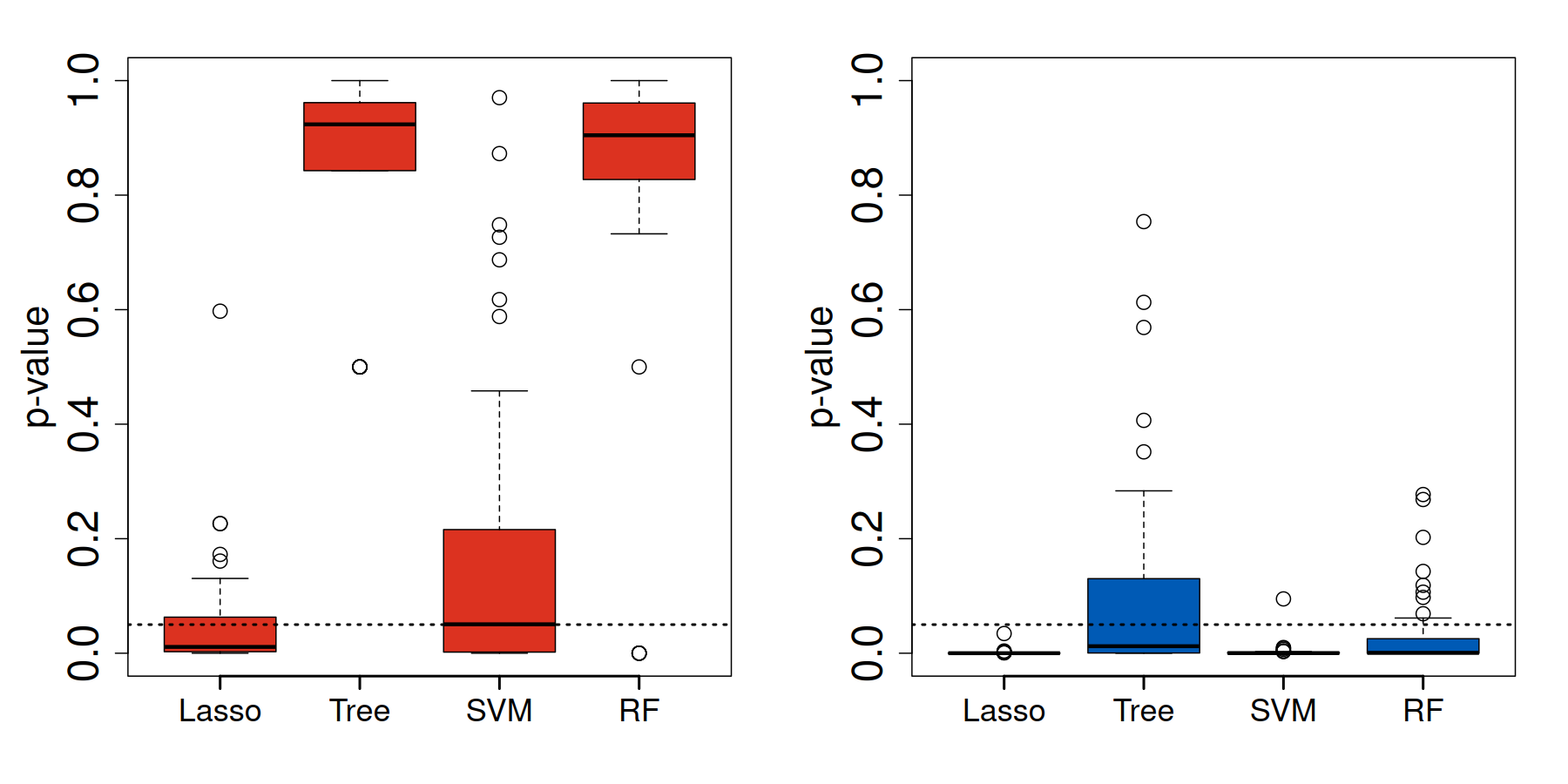}
\caption[Estimation of p-values for the simulations]{Simulated data: Estimation of the p-values for the statistical test with alternative hypothesis ``the risk of $g_i$ is greater than the risk of $g_{Linear}$", where $i\in\{\text{Lasso, Tree, SVM, RF}\}$, see Section \ref{sec::classif_descr} for the description of the classifiers. Cross-validation is performed with $50$ repetitions (within the same procedure as in the application), $70\%$ train and $30\%$ test, threshold $u$ is the $0.97$ quantile of $H$. In red (left) are the estimates when $v=0$ and in blue (right) for $v=0.4 u<u$. Significance level of $0.05$ is depicted in dotted line.}
\label{fig::p_values_sim}
\end{figure}

We apply the developed test to the simulation presented in Section \ref{sec::simulations}, comparing the linear classifier to the four other classifiers. The p-values are estimated using the same cross-validation procedure, testing whether the risk of each classifier exceeds the one of the linear classifier. As shown in Figure \ref{fig::p_values_sim}, for $v=0$, p-values are mostly below $0.05$ for logistic regression, above $0.05$ for decision tree and random forest, and mixed for SVM. This suggests that the risk of logistic regression classifier is higher than that of the linear classifier, while the risk associated with the other three classifiers appears to be lower. For $v>0$, p-values are mostly below $0.05$, indicating that the linear classifier has the lowest risk. When $v>0$, classifier comparisons improve, as seen in the right panel of Figure \ref{fig::p_values_sim}.

\begin{figure}
\centering
\includegraphics[scale=0.4]{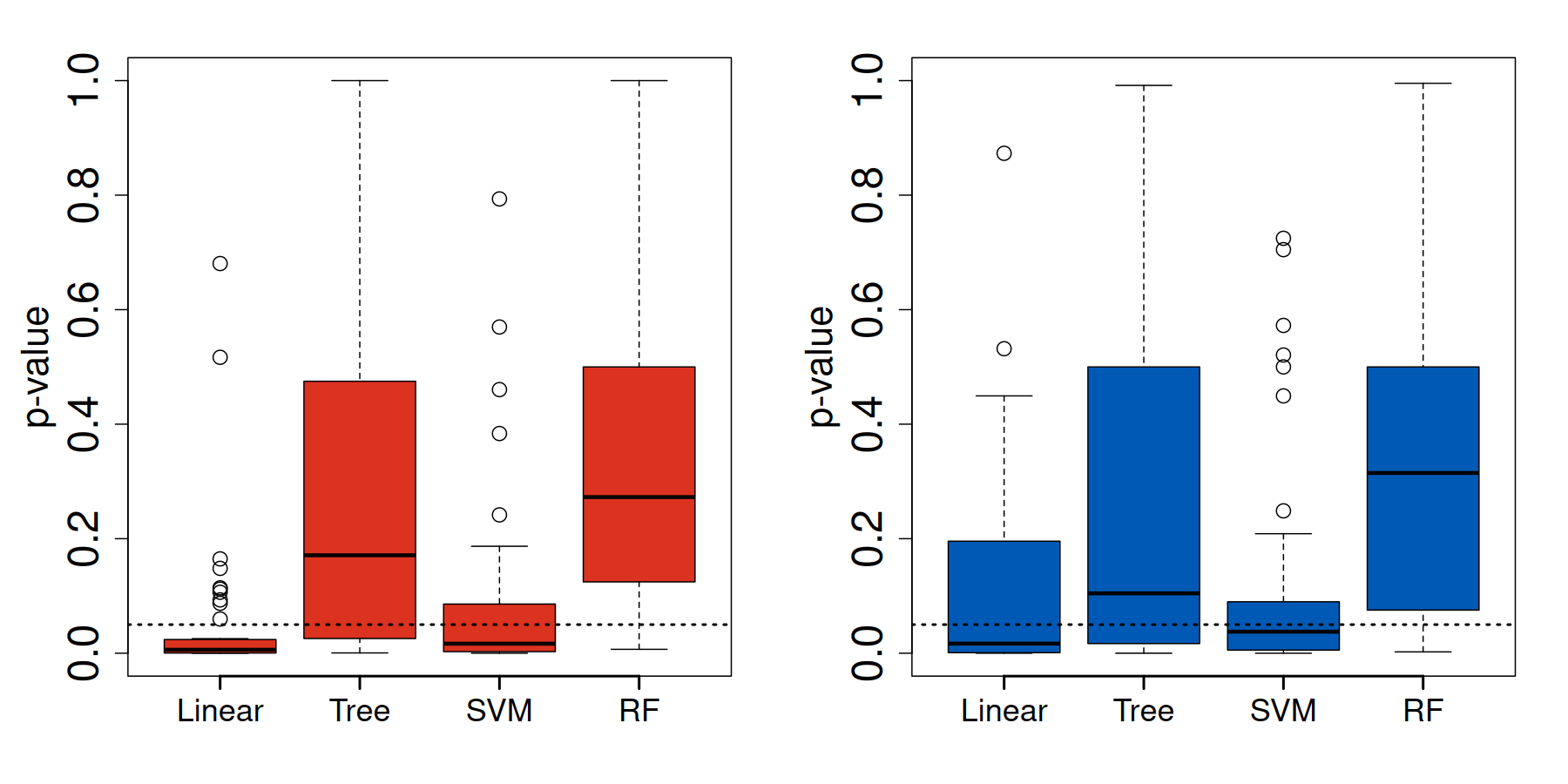}
\caption[Estimation of p-values for the Danube river discharges]{Application data (Danube river discharges): Estimation of the p-values for the statistical test with alternative hypothesis ``the risk of $g_i$ is greater than the risk of $g_{Lasso}$", where $i\in\{\text{Linear, Tree, SVM, RF}\}$. Same cross-validation procedure as in the simulation part is performed, threshold $u$ is the $0.85$ quantile of $X_1$. In red (left) are the estimates when $v=0$ and in blue (right) for $v=0.6 u<u$. Significance level of $0.05$ is depicted in dotted line.}
\label{fig::p_values_app}
\end{figure}

For the application, we test, with the same framework, whether the risk of each classifier exceeds the risk of the logistic regression classifier. As shown in Figure \ref{fig::p_values_app}, the logistic regression classifier appears to have a lower risk than linear and SVM classifiers but a comparable or higher risk than decision tree and random forest classifiers.

In both the simulation and the application, we obtain the results indicated in Figures \ref{fig::simuRisk} and \ref{fig::hrv_river2_res}.
\newpage

\section{Linear Classifiers}\label{sec:lin_class}
\subsection{Definition, Basic Properties and Inference}

In this section, we consider a specific type of classifiers which in this paper is referred to as linear classifiers, \textit{i.e.}\ classifiers of the form
$$ \gu_{\bm \theta}(\bm X) = \begin{cases}
  +1, & \quad \bm \theta^\top \bm X > u, \\
 -1, & \quad \bm \theta^\top \bm X \leq  u,
\end{cases} \qquad  \text{for some } \bm \theta \in [0,\infty)^d, $$
where, for notational convenience, we assume that $\bm X$ is nonnegative, i.e.\ $\bm X \in [0,\infty)^d$. Intuitively, these classifiers assume that each covariate -- provided that it has an effect at all --- has an amplification effect on the extremity of the event to be classified and model the strength of these effects by the weight vector $\bm \theta$.
To obtain an optimal linear classifier of $\gu_{\bm \theta}(\bm X)$, \textit{i.e.}\ some weight vector $\bm \theta^*$ such that the extremal risk $R(g_{\bm \theta^*})$ gets minimal, we need to impose some joint extremal dependence structure on $g$ and $Y$ {\color{black} which allow us to calculate $R$ from \eqref{eq:R-xi-c}.}

Even though some of the results can also be obtained in a similar manner in a more general framework for the extremal conditional risk $R_{\varepsilon}$, henceforth, we will focus on the asymptotically dependent case where we might find some optimal classifier with unconditional extremal risk $R(g_{\bm \theta^*}) < 1$. As discussed before, in this case, {\color{black} $g_{\bm \theta^*}$ has to be tail equivalent to $Y$, \textit{i.e.}\
\begin{equation} \label{eq:tail-linear}
 \frac{\PP((\bm \theta^*)^\top \bm X > u)}{\PP(\Yu = +1)} \to c(g) \in (0,\infty) \quad \text{as } u \to \infty.
\end{equation}
As the tail behavior of a linear combination is mainly driven by the component with the most heavy tail, it is natural to assume the tail behavior of $\|\bm X\|_\infty$ jointly with the asymptotic behavior of $u \mapsto \PP(\Yu=+1)$. In order to obtain asymptotic dependence, we need an additional assumption similar to joint regular variation in classical extreme value theory.}

{\color{black}
To establish such a condition, let us have a closer look at the function $u \mapsto \PP(\Yu=+1)$ and assume that it is eventually continuous. Then, we can define the random variable 
$$ H = \sup\{ u \geq 0: \, \Yu = +1\}.$$
By definition of $N$ and the fact that the sets $\{\Yu =+1\}$ are nested for for different levels $u$, we note that $H > u$ implies that $\Yu = +1$ while $H < u$ implies that $\Yu = -1$. To cover the case $H=u$, we note that
\begin{align*}
\PP(H=u) ={}& \PP( \Yv=+1 \text{ for all } v < u \text{ and } \Yv=-1 \text{ for all } v > u) \\
={}& \PP( \Yv=+1 \text{ for all } v < u) - \PP( \Yv=+1 \text{ for some } v > u) \\
={}& \lim_{v \uparrow v} \PP(\Yv = +1) - \lim_{v \downarrow v} \PP(\Yv = +1) = 0
\end{align*}
for sufficiently large $u$ as the function $v \mapsto \PP(\Yv=+1)$ as assumed to be eventually continuous. Thus, for sufficiently large $u$ and with probability one,
$$ \Yu = \begin{cases} 
                 +1, & \text{if } H > u \\
                 -1, & \text{if } H \leq u\\
                \end{cases}.
                 $$
Thus, Equation \eqref{eq:tail-linear} means that $(\bm \theta^*)^\top \bm X$ and $H$ are tail equivalence, while asymptotic dependence between $g$ and $Y$ 
is equivalence to asymptotic dependence of $(\bm \theta^*)^\top \bm X$ and $H$ in the classical notion of extreme value theory.
}

A natural assumption is therefore that $(\bm X,H)$ is jointly regularly varying
on $[0,\infty)$ with index $\alpha > 0$, \textit{i.e.}\ there exist an $\alpha$-Pareto random variable $P$ and, independently of $P$, a random vector 
$(\bm \Gamma, \Omega) \in [0,\infty)^{d} \times [0,\infty)$, the so-called \emph{spectral tail vector}, on the unit sphere $\{\bm x \in [0,\infty)^{d+1}: \,\|\bm x\|_\infty=1\}$ such that
$$ \left( \left(\frac{\|(\bm X,H)\|_\infty}u, \frac{(\bm X,H)}{\|(\bm X,H)\|_\infty}\right)  \, \bigg| \, \|(\bm X,H)\|_\infty > u \right) 
    \xrightarrow[]{d}( P,(\bm \Gamma,\Omega) ), \mbox{as }  u \to \infty.
    $$
Here, we additionally assume that % either the full random vector $\Gamma$ nor the random variable $\Omega$ vanish almost surely, i.e.
\ $\PP(\|\bm \Gamma\|_\infty > 0) > 0$ and
$\PP(\Omega > 0) > 0$.
In this setup, \ $\|\bm X\|_\infty$ and $H$ are tail equivalent in the sense that 
\begin{align*}
\lim_{u \to \infty} \frac{\PP( \|\bm X\|_\infty > u)}{\PP(H > u)}
   ={}& \lim_{u \to \infty} \frac{\PP(\|\bm X\|_\infty > u \mid \|(\bm X,H)\|_\infty > u)}
                              {\PP(H > u \mid \|(\bm X,H)\|_\infty > u)} \\
    ={}& \frac{\PP( P \cdot \|\bm \Gamma\|_\infty > 1)}{\PP( P \cdot \Omega > 1)} 
    = \frac{\EE(\|\bm \Gamma\|_\infty^\alpha)}{\EE(\Omega^\alpha)} \in (0,\infty),
\end{align*}
This is the minimal requirement on the link between the covariates $\bm X$ and the unobserved extremes of $H$
essentially saying that at least one component of $\bm X$ is tail equivalent to $H$. It is important to highlight  that we do not exclude the case that $\Gamma_i = 0$ a.s.\ for some $i \in \{1,\ldots,d\}$ which means that $X_i$ possesses a lighter tail than $H$. This property can be read off from the quantity
$$ c_i = \lim_{u \to \infty} \frac{\PP(X_i > u)}{\PP(H>u)} = \frac{\EE(\Gamma_i^\alpha)}{\EE(\Omega^\alpha)} \in [0,\infty).$$
Thus, $\Gamma_i = 0$ a.s.\ if and only if $c_i=0$. 
By including this case, we therefore admit that most of the components of $\bm X$ may not contribute to the extremes of the vector $H$.

Under these conditions, we obtain that, for all $\bm \theta \in [0,\infty)^d$, the classifier $g_{\bm \theta}$ is an extremal classifier as
\begin{align}\label{eq:r-lin-regvar}
R(g_{\bm \theta}) ={}& \lim_{u \to \infty} \frac{\PP[\max\{\bm \theta^\top \bm X, H\} > u] - \PP[\min\{\bm \theta^\top \bm X, H\} > u]}{\PP(H > u \text{ or } \bm\theta^\top \bm X > u)}\nonumber \\
={}& \frac{\EE\left(\max\{\bm \theta^\top \bm \Gamma, \Omega\}^\alpha\right) -  \EE\left(\min\{\bm \theta^\top \bm \Gamma, \Omega\}^\alpha \right)}{\EE\left(\max\{\bm \theta^\top \bm \Gamma, \Omega\}^\alpha\right) } {}={} 1 - \frac{\EE\left( \min\{\bm \theta^\top \bm \Gamma, \Omega\}^\alpha \right)}{\EE\left(\max\{\bm \theta^\top \bm \Gamma, \Omega\}^\alpha\right) }
 \in [0,1].
\end{align}

Equation \eqref{eq:r-lin-regvar}  implies that  the function $\bm \theta \mapsto 
R(g_{\bm \theta})$ is well-defined and continuous on $[0,\infty)^d$. Its value does not depend on those components $\theta_i$ for which $\Gamma_i=0$ a.s., which is equivalent to $c_i=0$ as discussed above.
Thus, in the following, we will consider this function only on the parameter set   
 $$ C = \{ \bm \theta \in [0,\infty)^d: \, \theta_i=0 \text{ for all $i$ s.t. } c_i = 0\}, $$
containing all the relevant information -- here, note that, in practice, identifying
 the components $i \in \{1,\ldots,d\}$ such that $c_i = 0$ a.s., from a given data set is a 
 necessary step for the correct specification of the set $C$.   

From the consideration in the Introduction, it can be easily seen that $R(g_{\bm 0})=1$; the case $\bm \theta=\bm 0$ {\color{black} corresponding} to the trivial always optimistic classifier. Furthermore, denoting the set of indices $j$ with $c_j > 0$ by $J$, we can see that
\begin{equation} \label{eq:r-limit}
R(g_{\bm \theta}) \geq 1 - \frac{\EE(\Omega^\alpha)}{\|\bm \theta\|^\alpha_\infty \cdot \min_{j \in J} \EE(\Gamma_j^\alpha)} \to 1 
\end{equation}
as $\|\bm \theta\|_\infty \to \infty$. By the continuity of $\bm \theta \mapsto R(g_{\bm \theta})$, we obtain that the function attains a global minimum on the domain $C$.

\begin{proposition}\label{proof: RV and unique mimizer}
Additionally to the assumptions above on the joint distribution of $(\bm X,H)$ with $\alpha > 1$, 
assume that there exists a function $a(u)$ with $a(u) \to 0$ as $u \to \infty$
such that
\begin{equation} \label{eq:uniform}
 \PP( u^{-1} \bm X \in A \mid \|(\bm X,H)\| > u) \leq (1 + a(u)) \PP( P \bm \Gamma \in A)
\end{equation}
for all $A \subset [0,\infty)$.
Furthermore, let $u_n \to \infty$ and $n \PP(H > u_n) \to \infty$ such that, for every compact subset $K \subset C$,
\begin{align}
    \sup_{\bm \theta \in K} \sqrt{n \PP(H > u_n)} \left| \frac{\PP({\color{black}\gun_{\bm \theta}(\bm X)} \neq \Yun)}{\PP(H > u_n)} - \frac{\EE\left(\max\{\bm \theta^\top \bm \Gamma, \Omega\}^\alpha\right) -  \EE\left(\min\{\bm \theta^\top \bm \Gamma, \Omega\}^\alpha \right)}{\EE(\Omega^\alpha)} \right| ={}& 0\label{eq:unbiased1} 
\end{align}
and
\begin{align}
    \sup_{\bm \theta \in K} \sqrt{n \PP(H > u_n)} \left| \frac{\PP(\max\{{\color{black}\gun_{\bm \theta}(\bm X)}, \Yun\}=1)}{\PP(H > u_n)} - \frac{\EE\left(\max\{\bm \theta^\top \bm \Gamma, \Omega\}^\alpha\right)}{\EE(\Omega^\alpha)} \right| ={}& 0. \label{eq:unbiased2} 
\end{align}
 If the function $\bm \theta \mapsto R(g_{\bm \theta})$ has a unique minimizer $\bm \theta^*$ in $C$, then 
the estimator 
$$ \hat{\bm \theta}_{n,u_n} = \mathop{\mathrm{argmin}}_{\bm \theta \in C} {\color{black}\hat{R}_n^{(u_n,0)}(g_{\bm \theta})}. $$
is consistent, \textit{i.e.}\ $\hat{\bm \theta}_{n,u_n} \to_p \bm \theta^*$.
\end{proposition}
Given the set $C$, this result provides a strategy to find the optimal $\bm \theta$, \textit{i.e.}, the best linear classifier. Determining the set $C$ requires the 
identification of the relevant features, \textit{i.e.}\ the index set $J$ such that $c_j > 0$ if and only if $j \in J$. This is discussed in more detail in the following subsection.

\subsection{Feature Selection} \label{sec:feature_selection}

The notion of sparsity quickly comes into play when doing classification. This is all the more true when one is only interested in the extremes. Among the whole data set, only a small proportion will truly contribute to the extremal behavior of the variable of interest. 
Here, we develop a method to identify the informative signals in terms of extremes among a large data set, assuming that  the covariates and the observations $\Yu$ are tail comparable.
For a comprehensive review of existing methods on sparsity and multivariate extremes we highly recommend the work of \cite{Engelke-Ivanovs-2020}.

As we have seen above, for linear classifiers, all the relevant features $X_i$ necessarily satisfy
$c_i > 0$. Thus, feature selection can be based on estimation of the $c_i$ which can be done according to the following proposition.

%\emph{Case $\varepsilon > 0$:} \\
%If the tail function of $H$ is regularly varying with index $\alpha_H$ and the tail function of $X_i$ is regularly varying with index $\alpha_i$, we obtain
%$$ c_{i,\varepsilon} = \varepsilon^{\alpha_i - \alpha_H}. $$
%Thus, $c_{i,\varepsilon}=0$ only if $X_i$ possesses light tails (assuming that $H$ is heavy-tailed).
%
%\bigskip
%
%We will now argue that, for partially linear classifiers those components $X_i$ for which we have $c_{i,\varepsilon}=0$ do not have any effect.
%\begin{lemma}
%\label{lem::c_i}
%Assume that the joint distribution of $(\gbar(X),H)$ satisfies the Ramos and Ledford model
%$$
%\PP[\overline{g}(X) > u, H > v ] = L(u,v) (\Fbar_g(u) v^{-\alpha_H})^{-1/2\eta}, \qquad u,v >0.
%$$
%Then, the following holds true:
%\begin{enumerate}
%    \item[1.] If 
%    $$\lim_{u \to \infty} \frac{\PP(\gbar(X) > u \mid \gbar(X) >  \varepsilon u)}{\PP(H > u \mid H > \varepsilon u)} = 0,$$
%    we have that $\Rbar_\varepsilon(g)=1$ for all $\varepsilon \in (0,1]$.
%    \item[2.] If $\Rbar_\varepsilon(g) < 1$ and $c_{i,\varepsilon} = 0$, then, for all $c > 0$. we have that
%    $$ \Rbar_\varepsilon(g) = \Rbar_\varepsilon(h_i(c)), $$
%    where $\overline{h}_i(c) = \gbar(X) + c X_i$.
%\end{enumerate}
%\end{lemma}

%\subsubsection{Inference}

%We distinguish between the two cases $\varepsilon=0$ and $\varepsilon>0$. First, we consider the estimation of $c_i \coloneqq c_{i,0}$.

\begin{proposition}
Assume that
$$ c_i = \lim_{u \to \infty} \frac{\PP(X_i > u)}{\PP(\Yu=+1)} $$
exists. If $u_n \to \infty$ and $n \PP(\Yu=+1) \to \infty$, then
$$ \frac{\sum_{j=1}^n \one\{X_{j,i} > u_n\}}{\sum_{j=1}^n \one\{\Yun_j =+1\}} \to_p c_i. $$
If, additionally, 
$$ \sqrt{n \PP(\Yun=+1)} \left( \frac{\PP(X_i > u_n)}{\PP(\Yun=+1)} - c_i \right) \to 0$$
and
$$ \chi_i^* = \lim_{u \to \infty} \PP(X_i > u \mid \Yu=+1) \in [0,1] $$
exists, then, we have
$$ \sqrt{n \PP(H > u_n)} \left( \frac{\sum_{j=1}^n \one\{X_{j,i} > u_n\}}{\sum_{j=1}^n \one\{\Yun = +1\}} - c_i\right) \xrightarrow[n\to\infty]{d} \mathcal{N}\left(0, c_i \cdot \left[1 - 2 \chi^*_i + c_i\right]\right).$$
\label{prop::hatc}
\end{proposition}

\section{River network}
\label{sec::appRiverDep}

\begin{figure}[h]
\centering
\includegraphics[scale=0.4]{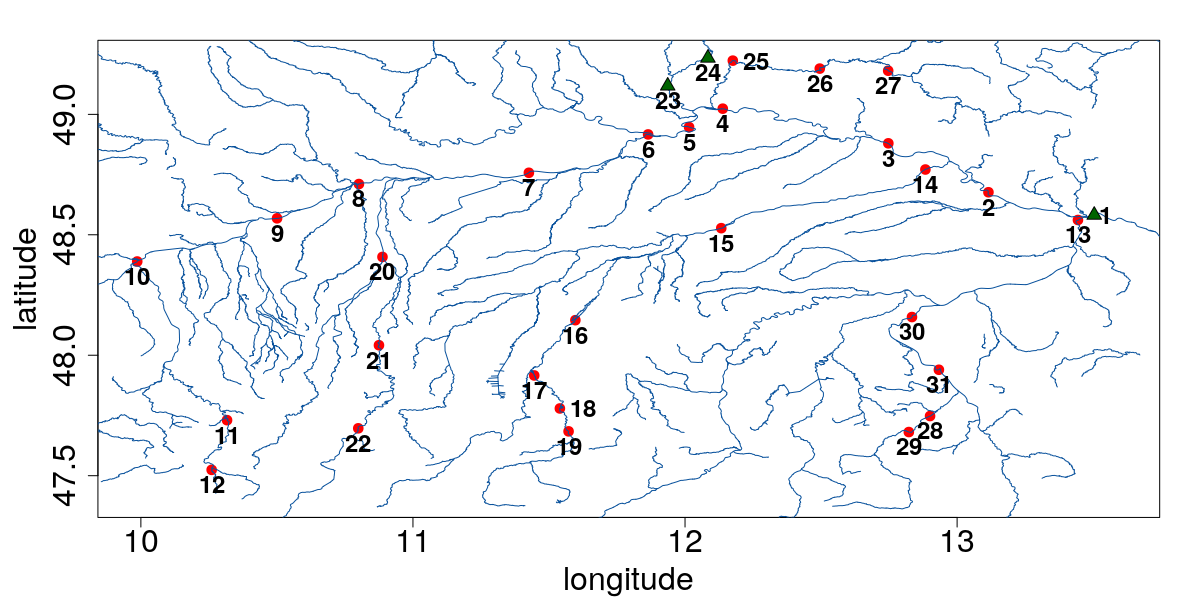}
\caption[River map of the upper Danube basin]{River map of the upper Danube basin, showing sites of the 31 gauging stations along the Danube and its tributaries. Water flows toward gauging station $1$. The stations represented by a green triangle shaped dot are the three stations of interest as described in Section \ref{sec: Danube river discharges}.}
\label{fig::Danube}
\end{figure}

This section deals with a simpler case than the application in Section \ref{sec: Danube river discharges}. Here an application could be the following: we want to know which stations should continue to be maintained to prevent extreme floods and maybe some stations are not necessary. 

Here we still want to predict the extreme events at station 1. We define an extreme event as an event exceeding the $0.85$ quantile of $X_1$ and we assume that the whole set of remaining stations is available. In this case, strong dependencies among station 1 and other stations can be observed. So, the main issue is to select
these stations following the procedure presented in Appendix \ref{sec:lin_class}.

\begin{figure}
\centering
\includegraphics[scale=0.45]{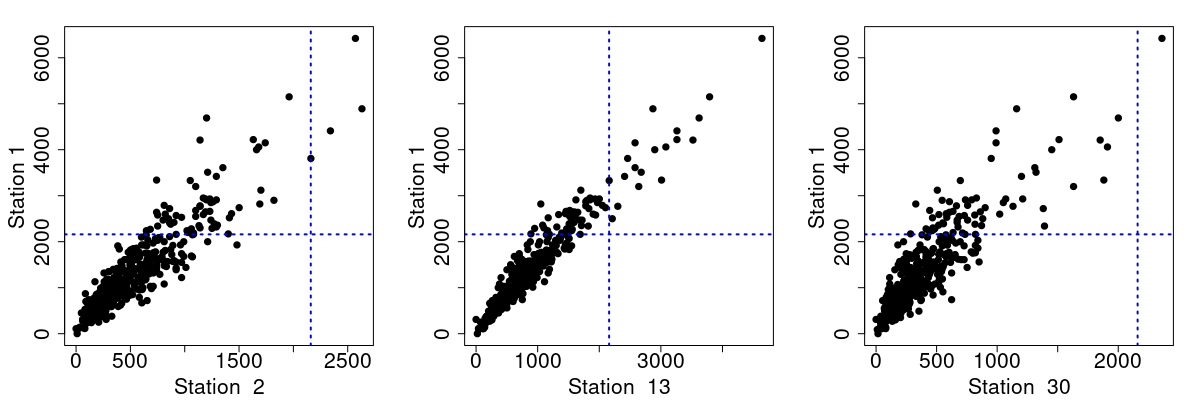}
\caption{Correlation plot between $X_1$ and the stations considered to contribute to the extremes of $X_1$ according to the table \ref{table::river} (\textit{i.e.} for which $c_i\neq 0$). The blue dotted lines represent the threshold $u$ defined by the $85$th percentile of $X_1$.}
\label{fig::scatter_river}
\end{figure}

As  detailed in Appendix \ref{sec:lin_class}, we identify the stations that may not contribute to the extremes of $X_1$ by estimating $(c_i)_{i\in\{2,\dots,31\}}$.
The estimation of the $c_i$ on all the data is presented in Table \ref{table::river}. We found that among the 30 stations, only three stations are relevant: stations 2, 13 and 30. Looking at Figure \ref{fig::Danube}, these stations correspond to the stations closest to $X_1$. Figure \ref{fig::scatter_river} shows the scatter plots between these stations and station 1.

% \begin{table}
% \caption{\label{table::river} Empirical estimates of $\hat{c}_i$ (as defined in proposition \ref{prop::hatc}) for each station. The values different from zero are highlighted in red.}
% \centering
% \begin{tabular}{cc}
%   \toprule
%  & $c_i$ \\ 
%   \midrule
%  X2 & \textcolor{red}{\textbf{0.06}} \\ 
%   X3 & 0.00 \\ 
%   X4 & 0.00 \\ 
%   X5 & 0.00 \\ 
%   X6 & 0.00 \\ 
%   X7 & 0.00 \\ 
%   X8 & 0.00 \\ 
%   X9 & 0.00 \\ 
%   X10 & 0.00 \\ 
%   X11 & 0.00 \\ 
%   X12 & 0.00 \\ 
%   X13 & \textcolor{red}{\textbf{0.30}} \\ 
%   X14 & 0.00 \\ 
%   X15 & 0.00 \\ 
%   X16 & 0.00 \\
%   \bottomrule
%   \end{tabular}
%   \hspace{2em} 
% \begin{tabular}{rr}
%   \toprule
%  & $c_i$ \\ 
%   \midrule 
%    X17 & 0.00 \\ 
%   X18 & 0.00 \\ 
%   X19 & 0.00 \\ 
%   X20 & 0.00 \\ 
%   X21 & 0.00 \\ 
%   X22 & 0.00 \\ 
%   X23 & 0.00 \\ 
%   X24 & 0.00 \\ 
%   X25 & 0.00 \\ 
%   X26 & 0.00 \\ 
%   X27 & 0.00 \\ 
%   X28 & 0.00 \\ 
%   X29 & 0.00 \\ 
%   X30 & \textcolor{red}{\textbf{0.02}} \\ 
%   X31 & 0.00 \\
%    \bottomrule
% \end{tabular}
% \end{table}

\begin{table}
\caption{\label{table::river} Empirical estimates of $\hat{c}_i$ (as defined in proposition \ref{prop::hatc}) for each station. The values different from zero are highlighted in red.}
\centering
\begin{tabular}{*{11}c}
  \toprule
 id & $X_2$ & $X_3$ & $X_4$ & $X_5$ & $X_6$ & $X_7$ & $X_8$ & $X_9$ & $X_{10}$ & $X_{11}$   \\
 $\hat c_i$ & \textcolor{red}{\textbf{0.06}} & 0.00 & 0.00 & 0.00& 0.00& 0.00& 0.00& 0.00& 0.00& 0.00\\
  \midrule
  id & $X_{12}$ & $X_{13}$ & $X_{14}$ & $X_{15}$ & $X_{16}$ & $X_{17}$ & $X_{18}$ & $X_{19}$ & $X_{20}$  & $X_{21}$ \\
  $\hat c_i$ &  0.00& \textcolor{red}{\textbf{0.30}}& 0.00& 0.00& 0.00& 0.00& 0.00& 0.00& 0.00& 0.00  \\
  \midrule 
  id & $X_{22}$& $X_{23}$& $X_{24}$& $X_{25}$& $X_{26}$& $X_{27}$& $X_{28}$& $X_{29}$ & $X_{30}$  & $X_{31}$ \\
  $\hat c_i$ & 0.00& 0.00& 0.00& 0.00& 0.00& 0.00& 0.00& 0.00& \textcolor{red}{\textbf{0.30}}& 0.00 \\
  \bottomrule
  \end{tabular}
\end{table}

\begin{figure}
\centering
\includegraphics[scale=0.4]{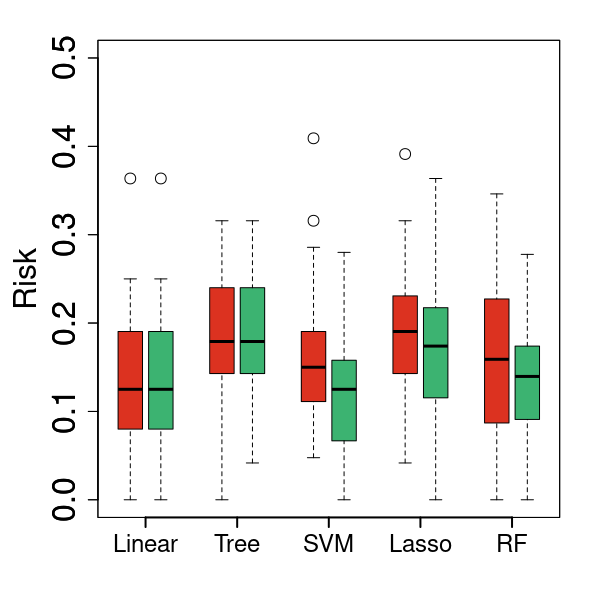}
\caption{Estimation of $R^{(u,0)}(g)$ for different classifiers $g$ (cross-validation - $70\%$ train, $30\%$ test - with $50$ repetitions). The red distributions come from the estimation with all the stations, and the green distributions represent the estimations with only the variables having $\hat{c}_i\neq 0$.}
\label{fig::comp_classif_dep}
\end{figure}

Once the contributing variables have been identified, we compare the performance of several classifiers, on the one hand keeping all the data and on the other hand keeping only the informative stations. Since there is a strong dependence between the data, we assume that it is sufficient here to look at the risk $R^{(u,0)}$. Comparison results are shown in Figure \ref{fig::comp_classif_dep}.

By definition of the linear classifier, the estimation is already done by keeping only the informative variables, which is why the estimates are identical for this specific classifier. As for the other classifiers, we see some improvements when keeping only the informative variables: the risk estimates are slightly smaller. This means that even if we remove a lot of information by going from 30 explanatory variables to 3, these 3 remaining stations contain all the information in terms of extremes of station 1.

\section{Proofs}\label{sec: proofs}

\subsection{Proof of Lemma \ref{lemma: sets}:}

Using that $\{ \gu(\bm X) = +1\} \subseteq \{\gv(\bm X) = +1\}$ and $\{\Yu=+1\}\subseteq\{\Yv=+1\}$ for $u \geq v$, we can write 
\begin{align*}
  & \PP( \gu(\bm X) = +1 \text{ or } \Yu=+1 \mid \gv(\bm X) = +1, \Yv=+1) \\
  ={}&  \frac{\PP( (\gu(\bm X) = +1 \text{ or } \Yu=+1), \gv(\bm X) = +1, \Yv=+1) }
  {\PP(  \gv(\bm X) = +1, \Yv=+1)},\\
 ={}& \frac{\PP( \gu(\bm X) = +1, \Yv=+1) + \PP( \gv(\bm X) = +1, \Yu=+1) -  \PP( \gu(\bm X) = +1, \Yu=+1)}
           {\PP( \gv(\bm X) = +1, \Yv=+1)}.
\end{align*}
In the same way, we have 
\begin{align*}
   & \PP(\gu(\bm X) \neq \Yu \mid \gv(\bm X) = +1, \Yv=+1) \\
={}& \frac{\PP(\gu(\bm X) = +1, \Yv=+1) + \PP( \gv(\bm X) = +1, \Yu=+1) - 2  \PP(\gu(\bm X) = +1, \Yu=+1)}
        {\PP(\gv(\bm X) = +1, \Yv=+1)}.
\end{align*}
Hence, we deduce that 
\begin{align*}
 & \Ruv(g) ={} \frac{\PP(\gu(\bm X) \neq \Yu \mid \gv(\bm X) = +1, \Yv=+1)}
                  {\PP(\gu(\bm X) = +1 \text{ or } \Yu=+1 \mid \gv(\bm X)=+1, \Yv=+1)} \\
={}& 1 - \frac{\PP(\gu(\bm X) = +1, \Yu=+1)}{\PP(\gu(\bm X) = +1, \Yv=+1) + \PP( \gv(\bm X) = +1, \Yu=+1) - \PP(\gu(\bm X) = +1, \Yu=+1)} \\   
={}& 1- \left[ \frac{\PP( \gu(\bm X) = +1, \Yv=+1)}{\PP(  \gu(\bm X) = +1, \Yu=+1)}  
    + \frac{\PP( \gv(\bm X) = +1, \Yu=+1)}{\PP(  \gu(\bm X) = +1, \Yu=+1)}-1 \right]^{-1}. 
\end{align*}
The expression given by \eqref{eq: sets} follows. 

Item (a) of the lemma follows from the representation \eqref{eq: sets} together with the inequality 
$$
\PP( U \mid V) \geq \PP( U \mid W) \mbox{, if the events $U$, $V$ and $W$ satisfy $U \subseteq V \subseteq W$.}
$$

For item (b), note that 
\begin{align*}
    & \Ruv(g) \leq  \Ruv(h) \\
    & \Longleftrightarrow  \left( 1 - \Ruv(g) \right)^{-1} \leq \left( 1 - \Ruv(h) \right)^{-1} \\
    & \Longleftrightarrow  \frac{\PP( \gu(\bm X) = +1 , \Yv=+1)}{\PP(  \gu(\bm X) = +1 , \Yu=+1)}  + \frac{\PP( \gv(\bm X) = +1 , \Yu=+1)}{\PP(  \gu(\bm X) = +1 , \Yu=+1)}  \\
    & \qquad \qquad \leq \frac{\PP(h^{(u)}(\bm X) = +1 , \Yv=+1)}{\PP( h^{(u)}(\bm X) = +1 , \Yu=+1)}  + \frac{\PP(h^{(v)}(\bm X) = +1 , \Yu=+1)}{\PP( h^{(u)}(\bm X) = +1 , \Yu=+1)},  \\
    & \Longleftrightarrow  \PP( \gu(\bm X) = +1 , \Yv=+1) + \PP( \gv(\bm X) = +1 , \Yu=+1) \\
    &  \qquad \qquad  \leq \frac{\PP(  \gu(\bm X) = +1 , \Yu=+1)}{\PP( h^{(u)}(\bm X) = +1 , \Yu=+1)}
                              \big[ \PP(h^{(u)}(\bm X) = +1 , \Yv=+1) + \PP(h^{(v)}(\bm X) = +1 , \Yu=+1)\big].
\end{align*}
As, by assumption, $\PP(\gu(\bm X) = +1, \Yu=+1) = \PP(h^{(u)}(\bm X) = +1, \Yu=+1) $   
and $\PP(\gu(\bm X) = +1, \Yv=+1) = \PP(h^{(u)}(\bm X) = +1, \Yv=+1) $, but also the inequality $\PP(\gv(\bm X) = +1, \Yu=+1) \leq \PP(h^{(v)}(\bm X) = +1, \Yu=+1)$ holds, we obtain
$$
\Ruv(g) \leq  \Ruv(h),
$$
\textit{i.e.}\ the second statement (b).

Item (c) is a direct consequence of  \eqref{eq: sets}.
\hfill $\square$

\subsection{Proof of Lemma \ref{lem: R(g)}:}

First, we note that, by definition, 
$$ \chi^*(g)  = c(g) \cdot \lim_{u \to \infty} \PP(  \Yu = +1 \mid \gu(\bm X) = +1) \leq c(g).$$
By taking the limits separately for every single term in \eqref{eq:ru-rewritten}, we directly obtain \eqref{eq:R-xi-c}. 

From that, we can immediately see that
$$ R(g)=0 \mbox{  if and only if }  1 + c(g) = 2 \chi^*(g). $$
Now, assume that $\chi^*(g) < c(g)$. Then,  $R(g)=0$ would imply that $1 + \chi^*(g) < 2 \chi^*(g)$ which is equivalent to $\chi^*(g) > 1$ in contradiction to the definition of $\chi^*(g)$. Thus, $R(g)=0$ implies $\chi^*(g) = c(g)$. Plugging this into $\eqref{eq:R-xi-c}$,
we obtain $c(g)=\chi^*(g)=1$. Conversely, $c(g)=\chi^*(g)=1$ obviously leads to $R(g)=0$. \hfill $\square$

\subsection{Proof of Lemma \ref{lem: R additive}:} 
%For technical reasons, the result will focus on the class of regular classifiers, i.e.\ classifiers $g$ such that the function $u \mapsto \PP(\gu=+1)$ is eventually continuous as $u \to \infty$.
%For a given regular classifier $g$, we can define the function
%$$ \gbar(\bm X) = \sup\{ u \geq 0: \, \gu(\bm X) = +1\}.$$
%By definition of $\gbar$ and the fact that the sets $\{\gu(\bm X) =+1\}$ are nested for for different levels $u$, we note that $\gbar(\bm X) > u$ implies that $\gu(\bm X) = +1$ while
%$\gbar(\bm X) < u$ implies that $\gu(\bm X) = -1$. To cover the case $\gbar(X)=u$, we note that
%\begin{align*}
%\PP(\gbar(\bm X)=u) ={}& \PP( \gv(\bm X)=+1 \text{ for all } v < u \text{ and } \gv(\bm X)=-1 \text{ for all } v > u) \\
%={}& \PP( \gv(\bm X)=+1 \text{ for all } v < u) - \PP( \gv(\bm X)=+1 \text{ for some } v > u) \\
%={}& \lim_{v \uparrow v} \PP(\gv = +1) - \lim_{v \downarrow v} \PP(\gv = +1) = 0
%\end{align*}
%for sufficiently large $u$ as the function $v \mapsto \PP(\gv(\bm X)=+1)$ as assumed to be eventually continuous for regular level-dependent classifiers. Thus, for sufficiently large $u$ and with probability $1$,
%$$ \gu(\bm X) = \begin{cases} 
%                  +1\text{, if } \gbar(X) > u \\
%                  -1\text{, if } \gbar(X) \leq u\\
%                \end{cases}.
%                  $$

{\color{black}
Let $(u_n)_{n \in \NN}$ be a sequence of positive thresholds with $u_n \to \infty$ as $u \to \infty$.
From the discussion at the beginning of Section \ref{subsec:extremalrisk} that we know that we can only get $R(g) < 1$, only if the sequence $\PP(\gun(\bm X)=+1) / \PP(\Yun = +1)$ is bounded away from $0$ and $\infty$, \textit{i.e.}\ there exist constants $0 < C_1 < C_2 < \infty$ such that
\begin{equation} \label{eq:tail-equiv}
 C_1 \leq \frac{\PP(\gun(\bm X)=+1)}{\PP(\Yun=+1)} \leq C_2 \quad \text{for all } n \in \NN.
\end{equation}
}
Thus, this will be assumed in the following.
 Let $w$ such that $\delta \coloneqq \lim_{u \to \infty} w(u)/u \in (0,1)$,  then we can write that
\begin{align*}
 & \PP(\Yun= + 1 \mid \gun(\bm X)=+1) =
\frac{\PP(f(\bm X) + N > u_n, \gun(\bm X) = +1)} {\PP(\gun(\bm X)=+1)} \\
\leq{}& \frac{\PP(f(\bm X)  > w(u_n), \gun(\bm X) = +1) }{\PP(\gun(\bm X)=+1)} + \frac{\PP(N > u_n-w(u_n), \gun(\bm X)=+1) }{\PP(\gun(\bm X)=+1)}.
\end{align*}

Since $\gun(\bm X)$ and $N$ are independent, the second term reduces to $\PP(N>u_n-w(u_n))$ which converges to $0$ as $n \to \infty$ since $u- w(u) \sim (1-\delta) u$ as $u$ gets large.

For the first term, we rewrite the ratio as follows
\begin{align*}
\frac{\PP(f(\bm X)  > w(u_n), \gun(\bm X)=+1) }{\PP(\gun(\bm X)=+1)} &\leq \frac{\PP(f(\bm X)  > w(u_n)) }{\PP(\gun(\bm X)=+1)} \\ 
& =  \frac{\PP(f(\bm X)  > w(u_n) )}{\PP(N > w(u_n))}  \frac{\PP(N  > w(u_n)) }{\PP(N>u_n)} \frac{\PP(N  > u_n) }{\PP(\gun(\bm X)=+1)}. 
\end{align*}

Since $w(u_n)\to \infty$ and $
\PP(f(\bm X) > u)= o\left( \PP(N > u) \right)$, the ratio $\frac{\PP(f(\bm X)  > w(u_n) )}{\PP(N >w(u_n))} $ goes to $0$ as $n \to \infty$.
From the assumption $w(u) \sim \delta u$  and regular variation of $N$, $\frac{\PP(N  > w(u_n)) }{\PP(N>u_n)}$ behaves as a constant when $n\to\infty$.
The only remaining term is $\frac{\PP(N  > u_n) }{\PP(\gun(\bm X) = +1)}$ which {\color{black}is bounded due to Equation \eqref{eq:tail-equiv}. So, 
 $$\lim_{n \to \infty}  \PP(\gun(\bm X) =+1 \mid \Yun = +1) =0.$$ By Equations \eqref{eq:ru-rewritten} and \eqref{eq:tail-equiv}, this implies
 $$ \lim_{n \to \infty} \Run(g) = 1.$$
 As this holds true for any sequence $(u_n)_{n \in \NN}$, we obtain $R(g)=1$.
 }
 \hfill $\square$
 
\subsection{Proof of Proposition \ref{prop: R eps}:}

The Ramos and Ledford model corresponds to the special of case of  \eqref{eq: mixing}:
For all $0 \leq v_1, v_2 \leq v$, we have
\begin{align*}
  \PP[g^{(v_1)}(\bm X) = + 1, 
       Y^{(v_2)} = +1]
    ={}& L(v_1, v_2) (v_1^{-\alpha_g} v_2^{-\alpha_Y})^{1/2\eta} \\
={}& c_{v_1,v_2} (\PP(g^{(v_1)}(\bm X) = +1))^a (\PP(Y^{(v_2)} = +1))^b
\end{align*}
with \begin{align*}
    \PP(g^{(v_1)}(\bm X) = +1) &=
    L_{g}(v_1) (v_1/u)^{-\alpha_g} u^{-\alpha_g}, \\
\PP(Y^{(v_2)} = +1) &= L_{Y}(v_2) (v_2/u)^{-\alpha_Y} u^{-\alpha_Y}
\end{align*}
for slowly varying functions $L_g$ and $L_Y$ 
and
$$
c_{v_1,v_2} = \frac{L(v_1,v_2)}{(L_g(v_1) L_Y(v_2))^{1/(2\eta)}}, \quad  
a = b = \frac{1}{2\eta}.
$$
Then, from Equation\eqref{eq: mixing cons},
$$
    \Ruv(g) 
=1- \left[ \frac{c_{v,u}}{c_{u,u}} (\PP( \gu(\bm X) = +1 \mid \gv(\bm X) = +1))^{-a}
    +  \frac{c_{u,v}}{c_{u,u}} (\PP( \Yu=+1 \mid \Yv=+1))^{-b} -1 
    \right]^{-1}.
$$
Letting $u$ get large and noting that,
by the continuity of $\ell^*$,
$$ \limeps \frac{c_{v,u}}{c_{u,u}} = \ell^*(\varepsilon,1) \quad \text{and} \quad
\limeps \frac{c_{u,v}}{c_{u,u}} = \ell^*(1,\varepsilon)
$$
provide the required result. \hfill $\square$.

\subsection{Proof of Propositions \ref{prop:emp_est} and \ref{prop:emp_est_biv}:}
\label{sec:appendixF5}

First of all, we note that Proposition \ref{prop:emp_est} is a special case of Proposition \ref{prop:emp_est_biv} with $g_1=g_2=g$ which implies $\cepsone=\cepstwo=1$. Thus, it suffices 
to show Proposition \ref{prop:emp_est_biv}. To this end, we define the constants
\begin{itemize}
\item $\displaystyle \rreps \coloneqq  \limeps \frac{\PP(\gu_1(\bm X) \neq \Yu, \gu_2(\bm X) \neq \Yu,
\gv_1(\bm X) =  \gv_2(\bm X) = \Yv=1)}{(p_{g_1} \wedge p_{g_2})(u_n,v_n)}, $
\item $\displaystyle \qepsonetwo \coloneqq  \limeps \frac{\PP(\gu_1(\bm X) \neq \Yu, \max\{\gu_2(\bm X), \Yu\}=1,
\gv_1(\bm X) = \gv_2(\bm X) = \Yv = 1)}{(p_{g_1} \wedge p_{g_2})(u_n,v_n)}, $
\item $\displaystyle \qepstwoone \coloneqq  \limeps \frac{\PP(\max\{\gu_1(\bm X), \Yu\} = 1, \gu_2(\bm X) \neq \Yu, \gv_1(\bm X) = \gv_2(\bm X) = \Yv = 1)}{(p_{g_1} \wedge p_{g_2})(u_n,v_n)}, $
\item $\displaystyle \peps \coloneqq \limeps \frac{\PP(\max\{\gu_1(\bm X), \Yu\} = 1, \max\{\gu_2(\bm X), \Yu\}=1, \gv_1(\bm X) = \gv_2(\bm X) = \Yv = 1)}{(p_{g_1} \wedge p_{g_2})(u_n,v_n)}, $
\end{itemize}
all of which exist in $[0,1]$ by the assumptions of the proposition.

For $n \in \NN$ and $i \in \{1,\ldots,n\}$, let
$$ \bm Z_{i,n} = \left( \begin{array}{c}
    \frac{\sqrt{\cepsone}}{\sqrt{n p_{g_1}(u_n,v_n)}} \Big[
    \one\{ \gun_1(\bm X_i) \neq \Yun_i, \gvn_1(\bm X_i)=\Yvn_i=1\} \\
    \qquad \qquad \qquad \quad - \PP(\gun_1(\bm X) \neq \Yun, \gvn_1(\bm X)=\Yvn=1) \Big] \\
    \frac{\sqrt{\cepsone}}{\sqrt{n p_{g_1}(u_n,v_n)}} \Big[
    \one\{ \max\{\gun_1(\bm X_i), \Yun_i\}=1, \gvn_1(\bm X_i)=\Yvn_i=1\} \\
    \qquad \qquad \qquad \quad - \PP(\max\{\gun_1(\bm X),\Yun\}=1, \gvn_1(\bm X)=\Yvn=1) \Big] \\
    \frac{\sqrt{\cepstwo}}{\sqrt{n p_{g_2}(u_n,v_n)}} \Big[
    \one\{ \gun_2(\bm X_i) \neq \Yun_i, \gvn_2(\bm X_i)=\Yvn_i=1\} \\
    \qquad \qquad \qquad \quad - \PP(\gun_2(\bm X) \neq \Yun, \gvn_2(\bm X)=\Yvn=1) \Big] \\
    \frac{\sqrt{\cepstwo}}{\sqrt{n p_{g_2}(u_n,v_n)}} \Big[
    \one\{ \max\{\gun_2(\bm X_i), \Yun_i\}=1, \gvn_2(\bm X_i)=\Yvn_i=1\} \\
    \qquad \qquad \qquad \quad - \PP(\max\{\gun_2(\bm X),\Yun\}=1, \gvn_2(\bm X)=\Yvn=1) \Big]
    \end{array} \right).  $$
First, we will  show that
\begin{align*}
 \sum_{i=1}^n \bm Z_{i,n}
                        \xrightarrow[n\to\infty]{d} {} &
\mathcal{N} \left( \bm 0, \bm \Sigma   \right),
\end{align*}
where
$$ \bm \Sigma = \left( \begin{array}{cccc} 
  \cepsone \Rbar_\varepsilon(g_1) & \cepsone \Rbar_\varepsilon(g_1)
  & \sqrt{\cepsone\cepstwo} \rreps & \sqrt{\cepsone\cepstwo} \qepsonetwo \\
  \cepsone \Rbar_\varepsilon(g_1) & \cepsone & \sqrt{\cepsone\cepstwo} \qepstwoone & \sqrt{\cepsone\cepstwo} \peps \\
  \sqrt{\cepsone\cepstwo} \rreps & \sqrt{\cepsone\cepstwo} \qepstwoone & 
  \cepstwo \Rbar_\varepsilon(g_2) & \cepstwo \Rbar_\varepsilon(g_2) \\
  \sqrt{\cepsone\cepstwo} \qepsonetwo  & \sqrt{\cepsone\cepstwo} \peps 
  & \cepstwo \Rbar_\varepsilon(g_2) & \cepstwo 
  \end{array} \right),$$
which, by the Cram\'er-Wold device is equivalent to
$$ \sum_{i=1}^n \bm a^\top \bm Z_{i,n}
                        \xrightarrow[n\to\infty]{d} {} 
\mathcal{N} \left( \bm 0, \bm a^\top \bm \Sigma \bm a \right) \quad \forall \bm a \in \RR^4, $$
To this end, we note that, by a straightforward calculation
$$ \lim_{n \to \infty} \sum_{i=1}^n \Var(\bm a^\top \bm Z_{i,n}) = \bm a^\top \bm \Sigma \bm a. $$
Furthermore, we have that
$$ |\bm a^\top \bm Z_{i,n}| \leq \|\bm a\|_1 \|\bm Z_{n,i}\|_\infty \leq 
\bm a\|_1 \cdot \left|  \frac{\sqrt{\cepsone}}{\sqrt{n p_{g_1}(u_n,v_n)}}  \vee  \frac{\sqrt{\cepstwo}}{\sqrt{n p_{g_2}(u_n,v_n)}} \right| \stackrel{n \to \infty}{\longrightarrow} 0. $$
Consequently, for each $\eta >0$, there exists $n_0 > 0$ such that
$ |\bm a^\top \bm Z_{i,n}| < \eta$ for all $n > n_0$ and, therefore,
$$ \sum_{i=1}^n \EE[ (\bm a^\top \bm Z_{i,n}|^2 \one\{ |\bm a^\top \bm Z_{i,n}|>\eta\}] = 0,$$
\textit{i.e.}\ the triangular array $(\bm a^\top \bm Z_{i,n})_{i=1,\ldots,n; n \in \NN}$ 
satisfies the Lindeberg condition. Thus, by Lindeberg's Central Limit Theorem
$$ \sum_{i=1}^n \bm a^\top \bm Z_{i,n}
                        \xrightarrow[n\to\infty]{d} {} 
\mathcal{N} \left( \bm 0, \bm a^\top \bm \Sigma \bm a \right) \quad \forall \bm a \in \RR^4. $$
This implies that
\begin{align*}
 \sqrt{n (p_{g_1} \wedge p_{g_2})(u_n,v_n)} &
\left( \begin{array}{c} \frac{\sum_{i=1}^n \one\{\gun_1(\bm X_i) \neq \Yun_i, \Yvn=\gvn_1(\bm X_i) = 1\}}{np_{g_1}(u_n,v_n)} - \frac{\PP(\gun_1(\bm X) \neq \Yun, \Yvn=\gvn_1(\bm X) = 1\}}{p_{g_1}(u_n,v_n)}\\
\frac{\sum_{i=1}^n \one\{\max\{\gun_1(\bm X_i), \Yun_i\}=1, \Yvn=\gvn_1(\bm X_i) = 1\}}{np_{g_1}(u_n,v_n)} - 1\\
\frac{\sum_{i=1}^n \one\{\gun_2(\bm X_i) \neq \Yun_i, \Yvn=\gvn_2(\bm X_i) = 1\}}{np_{g_2}(u_n,v_n)} - \frac{\PP(\gun_2(\bm X) \neq \Yun, \Yvn=\gvn_2(\bm X) = 1\}}{p_{g_2}(u_n,v_n)}\\
\frac{\sum_{i=1}^n \one\{\max\{\gun_2(\bm X_i), \Yun_i\}=1, \Yvn=\gvn_2(\bm X_i) = 1\}}{np_{g_2}(u_n,v_n)} - 1\\ \end{array} \right)  \\
={}& (1+ o(1)) \sum_{i=1}^n \bm Z_{i,n} \xrightarrow[n\to\infty]{d} {} 
\mathcal{N} \left( \bm 0, \bm \Sigma   \right),
\end{align*}
Applying the multivariate Delta method to the function
$$ (f_1, g_1, f_2, g_2) \mapsto \left( \frac{f_1}{g_1}, \frac{f_2}{g_2}\right)$$
yields the result. \hfill $\square$

\section{Proofs of the appendix}

\subsection{Proof of Proposition \ref{proof: RV and unique mimizer}} 

The proof is based on the following lemma which is proven in Subsection \ref{subsec:lem-unif}.

\begin{lemma} \label{lem:unif}
Under the assumptions from Proposition \ref{proof: RV and unique mimizer}, for every compact subset $K \subset C$, the sequences of processes $\{A_n(\bm \theta), \, \bm \theta \in K\}$ and $\{B_n(\bm \theta), \, \bm \theta \in K\}$ defined by
\begin{align*}
    A_n(\bm \theta) ={}& \sqrt{\frac{n}{\PP(H>u_n)}} \left( \frac 1 n  \sum_{i=1}^n \one\Big\{ \{\bm \theta^\top \bm X_i > u_n\} \triangle \{ H_i > u_n\} \Big\} - \PP( \gun_{\bm \theta}(\bm X)\neq Y^{(u_n)} ) \right) \\  
    B_n(\bm \theta) ={}& \sqrt{\frac{n}{\PP(H>u_n)}} \left( \frac 1 n  \sum_{i=1}^n \one\Big\{ \{\bm \theta^\top \bm X_i > u_n\} \cup \{ H_i > u_n\} \big\} - \PP( \max\{\gun_{\bm \theta}(\bm X), Y^{(u_n)}\}=1 ) \right) 
\end{align*}
 jointly converge to a centered bivariate Gaussian process $\{(A(\bm \theta), B(\bm \theta)), \ \bm \theta \in K\}$ weakly in $\ell^\infty(K)$.
\end{lemma}

 If the function $\bm \theta \mapsto R(g_{\bm \theta})$ has a unique minimizer $\bm \theta^*$, then, necessarily,  $R(g_{\bm \theta^*}) < 1$.\\
 Now, similarly to the notation above, let $J$ denote the set of indices $j$ with $c_j > 0$, and let
 us consider $\bm \theta \in C$ such that $\|\bm \theta\|_\infty > k_0$ for some constant $k_0 > 0$. Then,
 \begin{align*}
\widehat{R}_{n}^{(u_n,0)}(g_{\bm \theta})  ={}& 1 - \frac{\sum_{i=1}^{n} \one\{\min(\bm \theta^\top \bm X_i, H_i\} > u_n)\}} {\sum_{i=1}^{n} \one\{\max(\bm \theta^\top \bm X_i, H_i\} > u_n)\}}
 \geq 1 - \frac{\sum_{i=1}^{n} \one\{ H_i > u_n\}} {\min_{j \in J} \sum_{i=1}^{n} \one\{ k_0 X_{ij} > u_n\}} \\
 \stackrel{n \to \infty}{\longrightarrow}_p{}& 1 - \max_{j \in J} \frac{\EE(\Omega^\alpha)}{k_0^\alpha \EE(\Gamma_j^\alpha)} 
 \end{align*}
 where the right-hand side goes to $1$ as $k_0 \to \infty$. Thus, as $ \widehat{R}_{n}^{(u_n,0)}(g_{\bm \theta^*})\to_p 
 R(g_{\bm \theta^*}) < 1$, we obtain that, for sufficiently large $k_0 \gg \|\bm \theta^*\| $, with probability going to one,
 $$ \widehat{R}_{n}^{(u_n,0)}(g_{\bm \theta^*}) \leq \min_{\bm \theta \in C \setminus [0,k_0]^d} \widehat{R}_{n}^{(u_n,0)}(g_{\bm \theta})$$ 
 and, consequently, 
 \begin{align*}
	  \mbox{argmin}_{\bm \theta \in C} \widehat{R}_n^{(u_n,0)}(g_{\bm \theta})
	  ={}& \mbox{argmin}_{\bm \theta \in C \cap [0,k_0]^d} \widehat{R}_n^{(u_n,0)}(g_{\bm \theta}).
 \end{align*}
 Now, we note that, by Lemma \ref{lem:unif}, the bias conditions \eqref{eq:unbiased1} and \eqref{eq:unbiased2} and the functional delta method, $\widehat{R}_n^{(u_n,0)}(g_{\bm \theta})$ converges in probability to $R(g_{\bm \theta})$ uniformly on every compact subset of $C$. In particular,
 $$  \sup_{\bm \theta \in C \cap [0,K]^d} \left| \widehat{R}_n^{(u_n,0)}(g_{\bm \theta})
	    - R(g_{\bm \theta}) \right| \to_p 0. $$
 Thus,
  \begin{align*}
	  \mbox{argmin}_{\bm \theta \in C \cap [0,K]^d} \widehat{R}_n^{(u_n,0)}(g_{\bm \theta}) \to_p 
	  \mbox{argmin}_{\bm \theta \in C \cap [0,K]^d} R(g_{\bm \theta}) = \bm \theta^*.
 \end{align*}
\hfill $\square$

\subsection{Proof of Proposition \ref{prop::hatc}}

The proof runs analogously to the proof of Proposition \ref{prop:emp_est_biv}: For $n \in \NN$ and $j \in \{1,\ldots,n\}$, we define
$$ \bm Z_{j,n} = \left( \begin{array}{c}
    \frac{1}{\sqrt{n \PP(\Yun=+1)}} \Big[
    \one\{ X_{j,i} > u_n \} - \PP(X_i = +1) \Big] \\
    \frac{1}{\sqrt{n \PP(\Yun=+1)}} \Big[
    \one\{ \Yun_j = + 1\} - \PP(\Yun = +1) \Big] 
    \end{array} \right)  $$
in order to first show that
\begin{align*}
 \sum_{i=1}^n \bm Z_{i,n}
                        \stackrel{n \to \infty}{\longrightarrow}_d {}&
\mathcal{N} \left( \begin{pmatrix} 0 \\ 0\end{pmatrix}, \left( \begin{array}{cccc} 
  c_i & \chi_i^* \\
  \chi_i^* & 1
  \end{array} \right)   \right),
\end{align*}
which, by the Cram\'er-Wold device is equivalent to
\begin{equation} \label{eq:cramer-wold-ci}
\sum_{i=1}^n (a_1 Z_{i,n,1} + a_2 Z_{i,n,2})
                        \stackrel{n \to \infty}{\longrightarrow}_d {}
\mathcal{N} \left( \begin{pmatrix} 0 \\ 0\end{pmatrix}, \begin{pmatrix} a_1 & a_2 \end{pmatrix} \left( \begin{array}{cccc} 
  c_i & \chi_i^* \\
  \chi_i^* & 1
  \end{array} \right) \begin{pmatrix} a_1\\ a_2 \end{pmatrix} \right) \quad \forall a_1, a_2 \in \RR,
\end{equation}
A straightforward calculation leads to
$$ \lim_{n \to \infty} \sum_{i=1}^n \Var(a_1 Z_{i,n,1} + a_2 Z_{i,n,2}) = \begin{pmatrix} a_1 & a_2 \end{pmatrix} \left( \begin{array}{cccc} 
  c_i & \chi_i^* \\
  \chi_i^* & 1
  \end{array} \right) \begin{pmatrix} a_1\\ a_2 \end{pmatrix}. $$
By the inequality, we can see that
$$ |a_1 Z_{i,n,1} + a_2 Z_{i,n,2}| \leq  \frac{|a_1| + |a_2|}{\sqrt{n \PP(\Yun=+1)}} \stackrel{n \to \infty}{\longrightarrow} 0 $$
for each $\eta >0$, eventually $|\bm a^\top \bm Z_{i,n}| < \eta$ with probability one. Hence, we obtain
$$ \sum_{i=1}^n \EE[ |a_1 Z_{i,n,1} + a_2 Z_{i,n,2}|^2 \one\{ |a_1 Z_{i,n,1} + a_2 Z_{i,n,2}|>\eta\}] = 0,$$
\textit{i.e.}\ the triangular array $(\bm a^\top \bm Z_{i,n})_{i=1,\ldots,n; n \in \NN}$ 
satisfies the Lindeberg condition. Lindeberg's Central Limit Theorem then yields Equation \eqref{eq:cramer-wold-ci}.
The desired result follows by applying the Delta method to the function
$ (f,g) \mapsto  f/g$. \hfill $\square$

\subsection{Proof of Lemma \ref{lem:unif}}
\label{subsec:lem-unif}

We will proof the lemma by applying the Central Limit Theorem 2.11.9 in \cite{vdv-wellner-1996}.
To this end, we define the function spaces $\mathcal{A} = \{ a_{\bm \theta}, \ \bm \theta \in K\}$ and
$\mathcal{B} = \{ b_{\bm \theta}, \ \bm \theta \in K\}$ where
\begin{align*}
    a_{\bm \theta}: (0, \infty)^d \times (0, \infty)\to \{0,1\},& \quad 
    a_{\bm \theta}(\bm x,h) = \one\Big\{ \{\bm \theta^\top \bm x > 1\} \triangle \{ h > 1\} \big\} \\
    b_{\bm \theta}: (0, \infty)^d \times (0, \infty)\to \{0,1\},& \quad 
    b_{\bm \theta}(\bm x,h) = \one\Big\{ \{\bm \theta^\top \bm x > 1\} \cup \{ h > 1\} \big\}. 
\end{align*}
Then, with
$$ Z_{nl}(f) = \frac{1}{\sqrt{n \PP(H>u_n)}} f(u_n^{-1} \bm X_l, u_n^{-1} H_l), \quad f \in \mathcal{A} \cup \mathcal{B}, $$
for $l=1,\ldots,n,$ we have that
$$ \{A_n(\bm \theta), \ \bm \theta \in K\} = \left\{\sum\nolimits_{l=1}^n (Z_{nl}(f) - \EE Z_{nl}(f)), \ f \in \mathcal{A} \right\} $$
and 
$$ \{B_n(\bm \theta), \ \bm \theta \in K\} = \left\{\sum\nolimits_{l=1}^n (Z_{nl}(f) - \EE Z_{nl}(f)), \ f \in \mathcal{B} \right\}. $$
Now, we have that
$$ \max\left\{ \|Z_{nl}\|_{\mathcal{A}}, \|Z_{nl}\|_{\mathcal{B}} \right\} = \sup_{f \in \mathcal{A} \cup \mathcal{B}} |Z_{nl}(f)| \leq \frac 1 {\sqrt{n \PP(H>u_n)}} \ \text{a.s.} $$
for all $l=1,\ldots,n$ and $n \in \NN$.\\
Consequently, we check the Lindeberg condition: For $k \in \NN$, we have
$$ \lim_{n \to \infty} \sum_{l=1}^n \EE\left( \|Z_{nl}\|^k_{\mathcal{A} \cup \mathcal{B}} \one\{\|Z_{nl}\|_{\mathcal{A} \cup \mathcal{B}} > \eta \} \right) \leq
\lim_{n \to \infty} \frac n {\sqrt{n \PP(H>u_n)}^k} \one\{n \PP(H>u_n) < 1/\eta^2\} = 0 $$
as $n \PP(H>u_n) \to \infty$ by definition. For $k=2$, we obtain a Lindeberg type condition that ensures convergence of $A_n$ and $B_n$ to $A$ and $B$, respectively, in terms of finite-dimensional distributions.
For $k=1$, we obtain the Lindeberg type condition of Theorem 2.11.9 in \cite{vdv-wellner-1996}. \\

It remains to check the equi-continuity condition. In the following, to simplify notation, we assume that $C = [0,\infty)^d$. Then, for $\theta^{(1)}, \theta^{(2)} \in [a,b] \subset K \subset C$, we have that
$$ |a_{\theta^{(1)}}(u_n^{-1} \bm X,u_n^{-1} H) - a_{\theta^{(2)}}(u_n^{-1} \bm X,u_n^{-1} H)| \in \{0,1\} $$
and 
$$ |b_{\theta^{(1)}}(u_n^{-1} \bm X,u_n^{-1} H) - b_{\theta^{(2)}}(u_n^{-1}\bm X,u_n^{-1} H)| \in \{0,1\} $$
and the probability that any of those two expressions is equal to one is bounded by the probability
\begin{align*}
& \PP( \one\{ a^\top \bm X > u_n \} \neq\one\{ b^\top \bm X > u_n\})
={} \PP( a^\top \bm X \leq u_n, b^\top \bm X > u_n) \\
={}& \PP\left(\|(\bm X,H)\| > \frac{u_n}{\|K\| d}\right) \PP\left( a^\top \bm X \leq u_n, b^\top \bm X > u_n \, \Big| \, \|(\bm X,H)\| > \frac{u_n}{\|K\| d} \right) 
\end{align*}
where we use that 
$ b^\top \bm X > u_n$ implies that $\|X\| > u_n / (\|K\| d)$ with $\|K\| = \sup_{x \in K} \|x\|_\infty$.
Making use of the fact that $ \PP\left(\|(\bm X,H)\| > u_n / (\|K\| d)\right) \leq C_0 (\|K\|d)^\alpha \PP( H > u_n)$ for some constant $C_0 > 0$ and the bound given by Equation \eqref{eq:uniform}, we obtain that
\begin{align*}
& \PP( \one\{ a^\top \bm X > u_n \} \neq\one\{ b^\top \bm X > u_n\}) \\
\leq{}& C_0 (\|K\| d)^\alpha \PP(H > u_n) [1 + a(u_n/(\|K\|d))] 
\PP\left( P a^\top \bm \Gamma \leq \|K\|d, P b^\top \bm \Gamma > \|K\| d  \right) \\
={}& C_0 (\|K\| d)^\alpha \PP(H > u_n) [1 + a(u_n/(\|K\|d))] \EE_{\bm \Gamma}\left( \PP\left( P \in \left[  \frac{\|K\| d}{b^\top \bm \Gamma}, \frac{\|K\| d}{a^\top \bm \Gamma} \right] \right) \right) \\
\leq{}& C_0 \PP(H > u_n) [1 + a(u_n/(\|K\|d))] \EE\left( (b^\top \bm \Gamma)^\alpha - (a^\top \Gamma)^\alpha \right) \\
\leq{}& 2 C_0 \PP(H > u_n)  \|a-b\|
\end{align*}
provided that $u_n$ is sufficiently large 
as $a(u_n/(\|K\|d)) \to 0$.

Consequently,
$$ \sup_{\|f-g\| < \delta} \sum_{l=1}^n \EE\left[(Z_{nl}(f) - Z_{nl}(g))^2\right] = 2 C_0 \delta,$$ 
which tends to $0$ as $\delta \to 0$. From this inequality, it can also be seen that any partition of $K$ into hypercubes with length $\varepsilon^2/(2 C_0)$ leads to a valid $\varepsilon$-bracketing, \textit{i.e.}\ the number $N_\varepsilon \propto 1/\varepsilon^{2d}$
grows with a power rate and, so, $\sqrt{\log(N_\varepsilon)}$ is integrable. 

Thus, by Theorem 2.11.9, the the bivariate processes $(A_n,B_n)$ converge to a centered bivariate Gaussian process $(A,B)$ weakly in $\ell^\infty(K)$. The limiting covariance structure can be obtained from
\begin{align*}
    \Cov(A(\bm \theta_1), A(\bm \theta_2)) ={}& \lim_{n \to \infty}  \Cov(A_n(\bm \theta_1), A_n(\bm \theta_2)), \\
    \Cov(A(\bm \theta_1), B(\bm \theta_2)) ={}& \lim_{n \to \infty}  \Cov(A_n(\bm \theta_1), B_n(\bm \theta_2)), \\
    \Cov(B(\bm \theta_1), B(\bm \theta_2)) ={}& \lim_{n \to \infty}  \Cov(B_n(\bm \theta_1), B_n(\bm \theta_2)), \\
\end{align*}
for $\bm \theta_1, \bm \theta_2 \in K$.

\hfill  $\square$
\end{document}